\documentclass[]{interact}
\usepackage{epstopdf}
\usepackage{subcaption}
\usepackage{comment}
\usepackage[numbers,sort&compress]{natbib}
\usepackage{tabularx} 
\usepackage{multirow}
\usepackage{booktabs}
\usepackage{mathtools}
\newcolumntype{C}{>{\centering\arraybackslash}X} 
\bibpunct[, ]{[}{]}{,}{n}{,}{,}
\renewcommand\bibfont{\fontsize{10}{12}\selectfont}
\makeatletter
\def\NAT@def@citea{\def\@citea{\NAT@separator}}
\makeatother

\theoremstyle{plain}

\theoremstyle{definition}

\usepackage{color} 
\definecolor{OliveGreen}{rgb}{0,0.6,0}

\theoremstyle{remark}

 \newcommand{\sgn}{\mathop{\mathrm{sgn}}}
\DeclarePairedDelimiter\abs{\lvert}{\rvert}%
\newcommand\minipagetextwidth{0.68}

\begin{document}

\title{Estimation and Decomposition of Rack Force for Driving on Uneven Roads}

\author{
\name{Akshay Bhardwaj\textsuperscript{a}\thanks{CONTACT Akshay Bhardwaj Email: akshaybh@umich.edu}, Daniel Slavin\textsuperscript{b}, John Walsh\textsuperscript{b}, James Freudenberg \textsuperscript{c}, R. Brent Gillespie\textsuperscript{a}}
\affil{\textsuperscript{a}Department of Mechanical Engineering, University of Michigan, Ann Arbor, MI, USA;\\ \textsuperscript{b}Ford Motor Company, Dearborn, MI, USA;\\ \textsuperscript{c}Department of Electrical Engineering and Computer Science, University of Michigan, Ann Arbor, MI, USA}
}

\maketitle
\begin{abstract}
The force transmitted from the front tires to the steering rack of a vehicle, called the rack force, plays an important role in the function of electric power steering (EPS) systems. Estimates of rack force can be used by EPS to attenuate road feedback and reduce driver effort. Further, estimates of the components of rack force (arising, for example, due to steering angle and road profile) can be used to separately compensate for each component and thereby enhance steering feel. In this paper, we present three vehicle and tire model-based rack force estimators that utilize sensed steering angle and road profile to estimate total rack force and individual components of rack force. We test and compare the real-time performance of the estimators by performing driving experiments with non-aggressive and aggressive steering maneuvers on roads with low and high frequency profile variations. The results indicate that for aggressive maneuvers the estimators using non-linear tire models produce more accurate rack force estimates. Moreover, only the estimator that incorporates a semi-empirical Rigid Ring tire model is able to capture rack force variation for driving on a road with high frequency profile variation. Finally, we present results from a simulation study to validate the component-wise estimates of rack force.
\end{abstract}
\begin{keywords}
Rack Force Estimation; Steering Feel; Road Unevenness; Electric Power Steering; Rigid Ring Tire Model
\end{keywords}

\section{Introduction}\label{intro}

The torque experienced by a driver at the steering wheel, also referred to as steering feel, significantly influences a driver's perception of a vehicle \cite{strecker2014method, yang2014new, pick}. 
In modern cars, this torque feedback is primarily regulated by the Electric Power Steering (EPS) system \cite{gruner2008control}. An EPS system modulates the torque feedback by overlaying controlled amounts of torque on the steering column of the vehicle \cite{dornhege2017steering}. The objectives of the EPS system are to make the driving task easier, safer, and more comfortable while keeping the driver aware of road conditions \cite{yang2014new, dornhege2017steering, greul2012determining}. 

To achieve these objectives, the EPS system uses an estimate of rack force \cite{gruner2008control, dornhege2017steering, greul2012determining, fankem2014model}. Rack force is defined as the force transmitted from the front tires to the steering rack of a vehicle through the tie rods. 
Tire forces and moments, and hence the rack force, arise from the interaction of tires with the road. Naturally, rack force depends on the road profile, but also on how the road profile is traversed, and thus depends on the steering angle in combination with the road profile. When a driver performs a steering maneuver, the tire forces and moments and hence the rack force generally oppose the effort applied by the driver. The counteracting rack force increases the driver effort needed to steer the vehicle, however it also informs the driver of the vehicle state and the road conditions. Accordingly, EPS uses rack force estimates to attenuate the rack force and assist the driver in performing the maneuver, while leaving a portion of the counteracting force unattenuated to maintain driver awareness \cite{vinattieri2016target, greul2016method, strecker2014method, yang2014new, kezobo2014electric}. 

Apart from the EPS assist torque, rack force estimates are also used to determine the EPS torque needed to reject disturbances arising from elements internal to the steering system \cite{pick,dornhege2017steering, vinattieri2016target}.  
Lane keeping and steer-by-wire systems also utilize the estimates of rack force \cite{fankem2014model,nehaoua2012rack, bajcinca2006road, bhardwaj2019estimating}. Unfortunately, it is expensive to install reliable measurement systems for rack force in commercial vehicles \cite{yang2014new, bajcinca2006road}. As a result, estimation of rack force using real-time capable techniques has attracted the attention of researchers both in industry and academia \cite{strecker2014method,bajcinca2006road, dornhege2017steering, pick, fankem2014model, blommer2012systems, nehaoua2012rack, wang2016epas}.

One real-time capable technique used for rack force estimation utilizes system identification (SID) methods. An SID-based estimator uses data generated through driving experiments to identify a model between the measured output rack force and a measured input signal (such as rack displacement) \cite{wang2016epas,blommer2012systems}. Such estimators are computationally inexpensive and can be used in vehicles of different configurations \cite{wang2016epas}. However, current SID-based estimators can only estimate rack force due to the steering angle and ignore the effect of road profile variations on rack force.

\begin{figure}[h!]
	\centering
\begin{minipage}[b]{0.475\textwidth}
              \includegraphics[width=\textwidth]{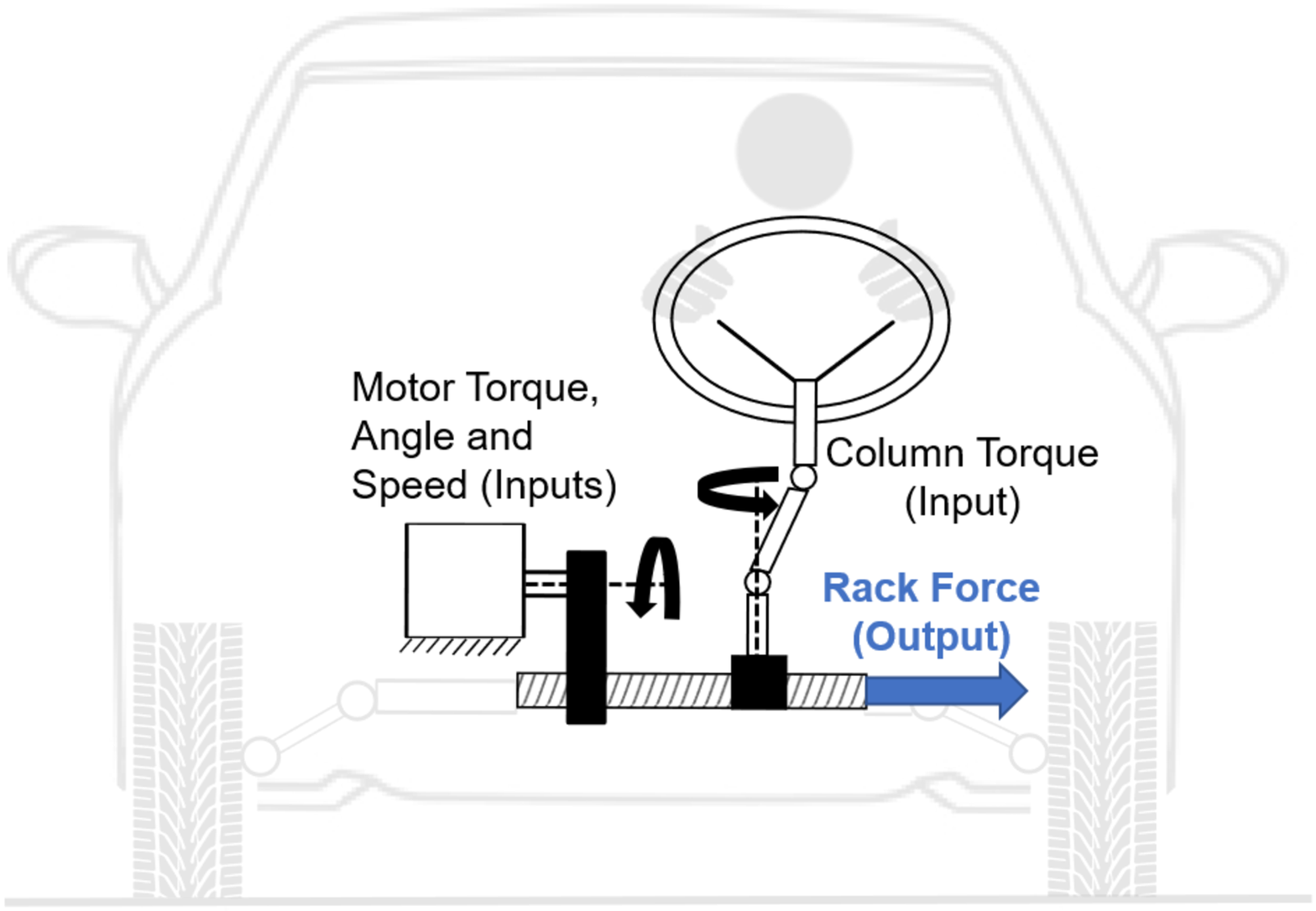}
				\vspace{-0.19cm}\subcaption{}
				\label{1aa}
\end{minipage}
        \begin{minipage}[b]{0.51\textwidth}
		\includegraphics[width=\textwidth]{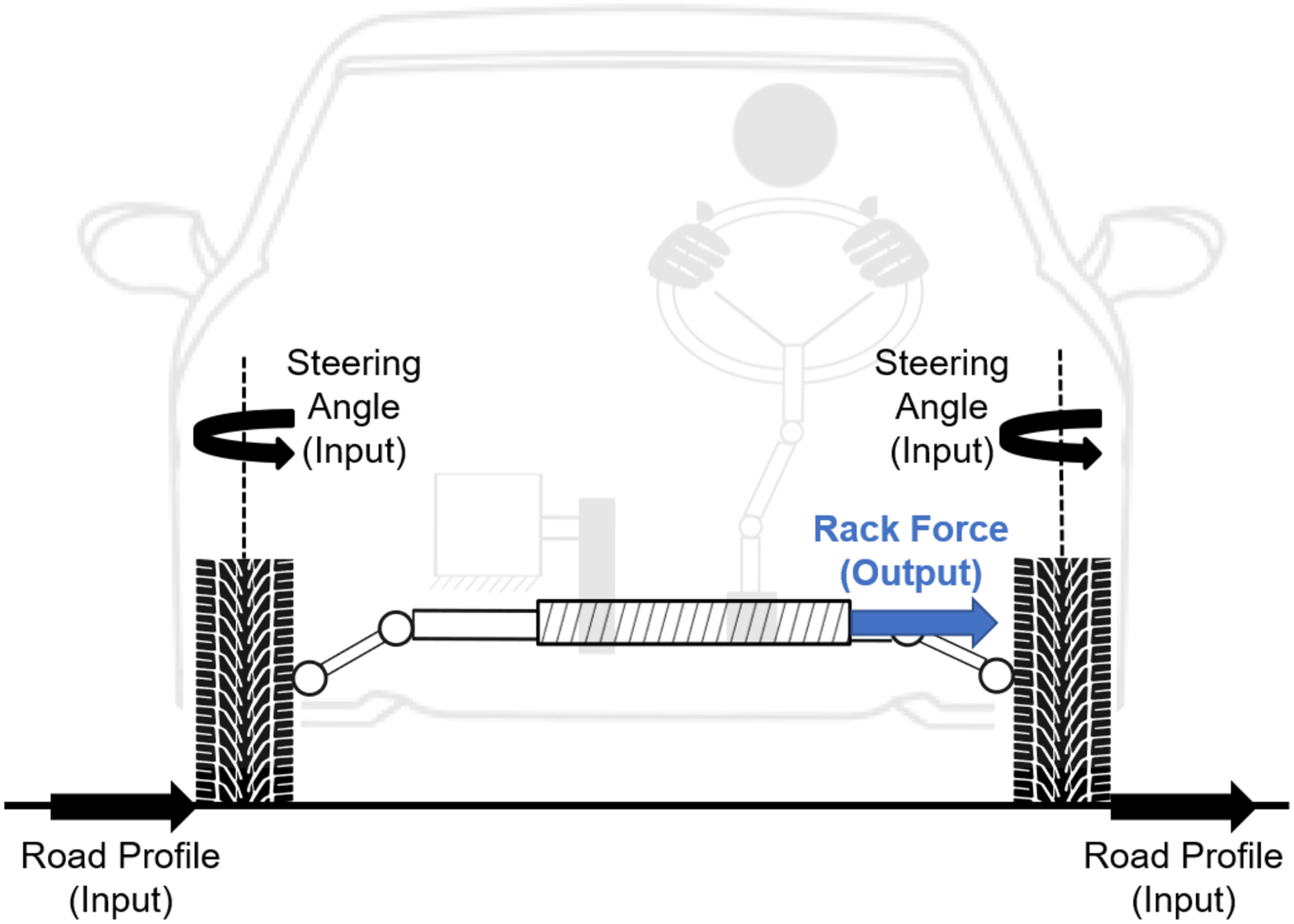}
		\subcaption{}
		\label{1bb}
\end{minipage}\hfill
	\caption{Two most common methods to estimate rack force due to steering angle and road profile: (a) In a steering model-based (SM-based) rack force estimator, the sensed EPAS motor angular position, speed, and torque along with the steering column torque are fed into an input observer to compute rack force. (b) In a vehicle and tire model-based (VTM-based) rack force estimator, the road profile and the steering angle are fed into a combined vehicle and tire model to compute rack force.} 
	\label{Fig:Themnew}
\end{figure}

A rack force estimator that disregards the effect of road profile variation can negatively affect the driver's safety. Studies show that any inability to account for road profile variation, such as road bank, road grade, and side-slopes, can result in long periods of unaccounted steering disturbances which might increase the chance of rollover and loss of steering control \cite{pick, tseng2001dynamic, eric2007estimation, yang2014new, viner1995rollovers, peters2006modeling}. Therefore, estimators that can 
account for road profile variation in rack force estimation have the potential to improve both the safety and comfort of the driver. Estimators shown in Fig. \ref{Fig:Themnew} utilize models of the steering and vehicle systems to estimate the effect of road profile variation on rack force.  
The estimator shown in Fig. \ref{Fig:Themnew}(a) is called a steering model-based (SM-based) rack force estimator, and the estimator shown in Fig. \ref{Fig:Themnew}(b) is called a vehicle and tire model-based (VTM-based) rack force estimator.

The SM-based rack force estimator uses a lumped parameter model of the steering system along with the EPS motor torque, position, and speed, and the steering column torque to produce an estimate of rack force \cite{dornhege2017steering, fankem2014model, blommer2012systems, nehaoua2012rack, weiskircher2015rack}. 
The SM-based rack force estimators have been widely used in EPS applications because of their ability to produce sensor-level-accurate rack force estimates. 
However, the SM-based estimators cannot estimate the contribution of road profile variation to rack force independent of the contribution of steering angle. 

An estimator that produces estimates of rack force due to road profile independent of the steering angle holds additional advantage for EPS control; two separately estimated components of rack force can be compensated individually to different degrees to enhance the steering feel. Several researchers have adopted the idea of performing decomposition of steering signals such as rack force \cite{strecker2014method, vinattieri2016target}, steering angle \cite{hackl2000method, karnopp1993motor}, and steering torque \cite{greul2012determining, yang2014new, ikeda2011electric, kezobo2014electric} with the aim of performing targeted compensation on the signal components and improving the steering feel. Currently, only the VTM-based rack force estimators (depicted in Fig. \ref{Fig:Themnew}(b)) are capable of producing component-wise estimates of rack force. A VTM-based estimator uses sensed steering angle and road profile together with a vehicle model and tire model to produce its estimate of rack force. 

VTM-based rack force estimators have appeared in various forms. Software packages such as CarSim and CarMaker use relatively more complex VTM-based estimators to produce highly accurate rack force estimates \cite{honisch2015improvement,toyohira2010validity}. A disadvantage of the estimators used in these packages is that they are computationally heavy and therefore cannot be used in real-time in production vehicles. They can, however, be used for running simulation studies to verify the estimation performance of other estimators \cite{marino2010asymptotic, gadola2014development}. Real-time capable VTM-based estimators appearing in the literature use simpler vehicle and tire models for rack force estimation and are computationally inexpensive. Most conventional VTM-based estimators ignore the presence of road profile variations and only consider the steering angle as an input when estimating the rack force (perhaps due to unavailability of real-time road profile measurements) \cite{dornhege2017steering,pick,koch2010untersuchungen,segawa2006preliminary}. Only two previous papers \cite{bhardwaj2019estimating, bhardwaj2020rack} introduced VTM-based rack force estimators that could incorporate road profile measured using sensors mounted in the vehicle. 

In this paper, we develop three VTM-based estimators that can estimate rack force using sensed steering angle and road profile inputs. We develop the estimators with the same vehicle model but three different tire models to isolate the effect of the tire model on estimation accuracy.
We present the results from three driving experiments to comment on which features of our tire models improve or reduce the accuracy of rack force estimation. In addition, we present a simulation study to test whether the estimates of rack force due to steering angle and rack force due to road profile produced by one of our estimators can potentially be used to perform targeted compensation.

The paper is organized as follows. In Section \ref{sec:mod_fram} we briefly discuss the overall structure of a VTM-based rack force estimator. Section \ref{sec:modeling} presents the details of the vehicle model and the three tire models used to estimate the rack force. In Section \ref{sec:methods} we describe the driving experiments which were used to compare the model fidelity of the three estimators, and the simulation setup that was used to produce the component-wise estimates of rack force. Section \ref{sec:results} presents the results and discussion for the driving experiments and the simulation study followed by Section \ref{sec:conclusions} that presents the conclusions of the paper.

\section{Modeling Framework}\label{sec:mod_fram}

\begin{figure}[h!]
\vspace{0 cm }
\begin{center}
\includegraphics[width=0.7\textwidth]{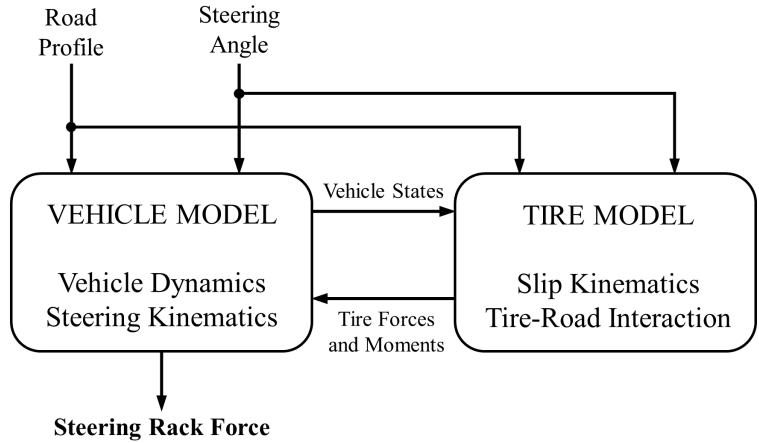}
\end{center}
\caption{Simplified structure of the VTM-based rack force estimator. The road profile and the steering angle are inputs to a vehicle model and tire model that in combination enable rack force estimation. The tire model includes a slip kinematics model to estimate the tire slip angles and a tire-road interaction model to estimate tire forces and moments. The vehicle model includes a vehicle dynamics model to estimate vehicle states and a steering kinematics model to estimate rack force.}
\label{Fig:overall3}
\end{figure}

Fig. \ref{Fig:overall3} shows the simplified structure of a VTM-based rack force estimator. For a relatively constant non-zero vehicle speed, the two inputs to the estimator are steering angle and road profile. While the steering angle is primarily governed by the driver, the road profile is determined by the environment. The most common road profile inputs are longitudinal slope, lateral slope, and cleat/pothole dimensions. The road profile can be sensed using advanced radar and LiDAR sensors or can be measured beforehand. Certain road profile inputs, especially road slopes, can also be estimated on the fly using on-board sensors (see \cite{bhardwaj2019estimating, pastor1995method}, for example).

In a VTM-based rack force estimator, the steering angle and road profile signals are first made available to a vehicle model that generates vehicle states. The same signals along with the vehicle states are then used in a tire model to obtain tire slip angles and tire forces and aligning moments. These forces and moments are in turn used in the vehicle model to generate vehicle states for the next time instant. Meanwhile, the tire aligning moments are used to determine the steering rack force using a steering kinematics model that relates tire aligning moments to the rack force. 

\section{Modeling}\label{sec:modeling}
In this section we present three VTM-based estimators that we developed to determine rack force. The three estimators are only distinguished by their tire models. The first estimator has a Linear Tire model \cite{rajamani2011vehicle}, and is called the LT Model. The second estimator has a nonlinear Brush Tire model (also known as the elastic foundation model) \cite{pacejka2005tire}, and is called the BT Model. The third estimator has a Rigid Ring tire model \cite{schmeitz2004semi}, and is called the RR Model. All the estimators are based on the same vehicle dynamics given by the 2DOF bicycle model presented in \cite{bhardwaj2019estimating}. 

The following assumptions apply to all rack force estimators presented in the paper:
\begin{enumerate}
    \item The tire parameters such as tire stiffness and damping and tire radius were assumed constant.
    \item Tire inertia and wheel camber were assumed negligible. Tire-road friction $\mu$ was assumed constant: $\mu = 1$.
    \item The effects of the vehicle's roll, pitch, acceleration and braking on vehicle states, normal tire force, and on steering angle were ignored. 
    \item The components of the steering system were assumed mass-less. The vehicle's mass was assumed constant.
    \item The influence of the suspension system on rack force was assumed negligible.
\end{enumerate}

In the following subsections, we first describe the vehicle model common to the three estimators followed by a description of the three tire models that distinguish the estimators. We then present how rack force was estimated using the tire models and briefly describe the assembly of the vehicle and tire models that enable rack force estimation. We conclude the section by describing how VTM-based estimators can be used to produce component-wise estimates of rack force and how the component-wise estimates can be used to perform targeted compensation to improve steering feel. 

\subsection{Vehicle Model}\label{vd}

Consider a vehicle of mass $m$ and yaw inertia $I$ driven with steering angle $\delta$ and speed $u$. Let the the vehicle yaw angle be $\psi$, and let the forces on the vehicle's front ($f$) and rear ($r$) tires in the longitudinal ($x$) and lateral ($y$) directions respectively be denoted by $F_{xf},F_{xr}$ and $F_{yf},F_{yr}$ (see Fig. 3). Likewise, let the tire aligning moments for the front and rear tires be denoted by $M_{zf}$ and $M_{zr}$. Then the two degrees of freedom of the vehicle, namely, lateral speed $v$ and yaw rate $\dot{\psi}$, or the lateral dynamics of the vehicle for driving on a flat road are governed by the following differential equations \cite{bhardwaj2019estimating}

 \begin{figure}[h!]
  \vspace{0 cm }
\begin{center}
 \includegraphics[width=0.5\textwidth]{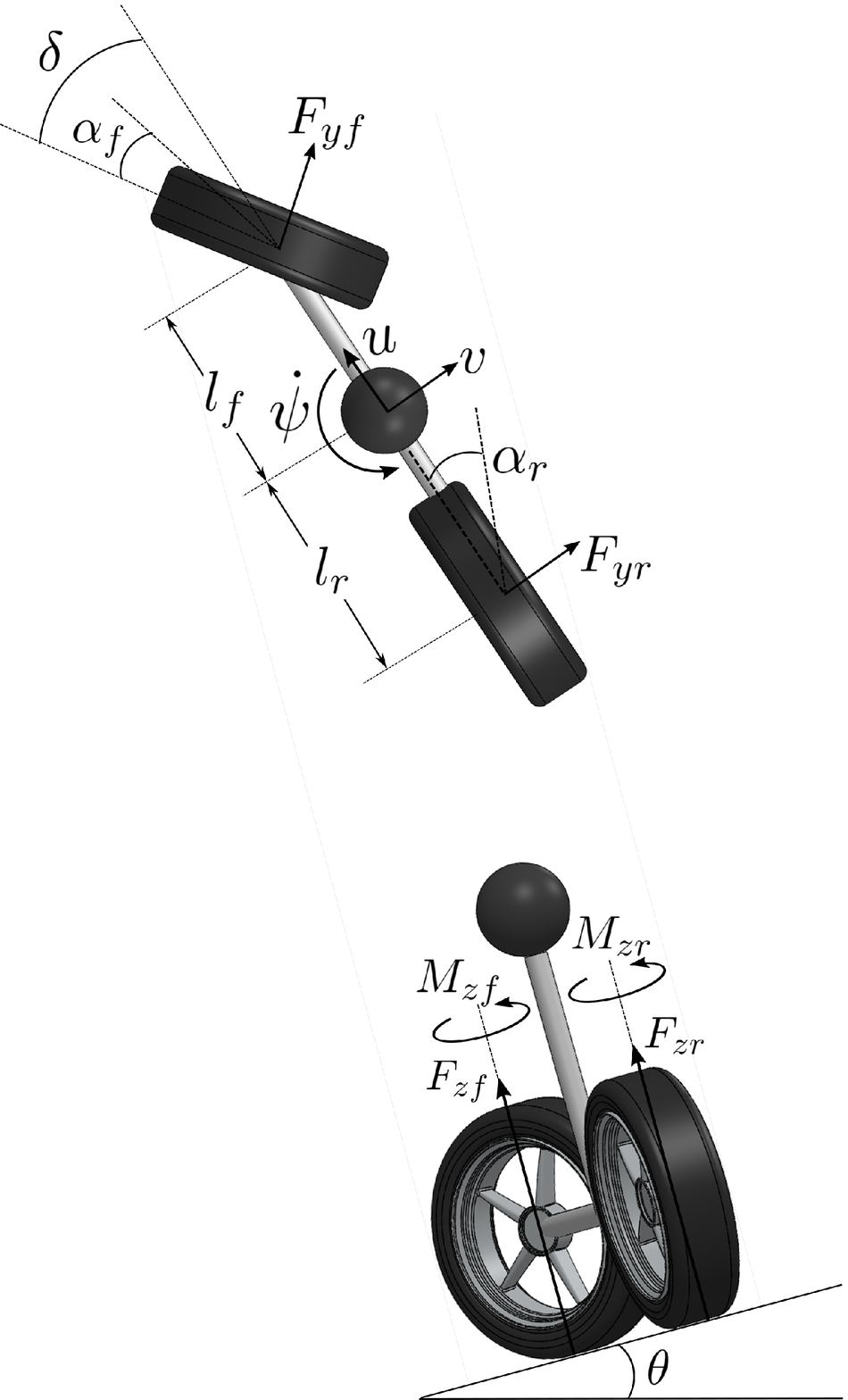}
 \end{center}
 \vspace{-.4 cm }
  \caption{Schematic of a 2DOF bicycle model. In the configuration shown, the bicycle drives along a road with lateral slope $\theta$ with speed $u$ and steering angle $\delta$ . The two degrees of freedom are lateral speed $v$ and yaw rate $\dot{\psi}$. The vertical tire forces, slip angles, tire lateral forces, and tire aligning moments are denoted by $F_{zi}$, $\alpha_i$, $F_{yi}$, and $M_{zi}$, respectively, where $i\in \lbrace f,r\rbrace$ denotes the front and rear tires.}
  \label{model}
\end{figure}

\begin{gather}\label{dyn_old}
\begin{aligned}
m\dot{v}+ mu\dot\psi = F_{xf}\sin\delta+F_{yf}\cos\delta+F_{yr}\\
{I}\ddot{\psi} = l_fF_{xf}\sin\delta+l_fF_{yf}\cos\delta-l_rF_{yr},
\end{aligned}
\end{gather}

\noindent where the distance between the center of mass of the vehicle and the centroid of the front tire contact patch is $l_f$ and the distance between the center of mass of the vehicle and the centroid of the rear tire contact patch is $l_r$.  

Assuming the steering angle $\delta$ remains small, the differential equations governing the lateral dynamics of the vehicle can be rewritten as 
\begin{gather}\label{dyn_1}
\begin{aligned}
m\dot{v}+ mu\dot\psi = F_{yf}+F_{yr}\\
{I}\ddot{\psi} =l_fF_{yf}-l_rF_{yr}.
\end{aligned}
\end{gather}

Now consider driving on an uneven road. Equation (\ref{dyn_1}) also applies for driving on a longitudinal slope or a road grade because the longitudinal slope does not significantly influence the lateral dynamics of the vehicle \cite{bhardwaj2019estimating}. Likewise, although driving over a cleat or a pothole changes the tire dynamics considerably, cleats or potholes do not directly influence the vehicle's lateral dynamics. On the other hand, for driving on a lateral road slope or a road bank (simply referred to as road slope in this paper), the differential equation for the lateral speed is different. If the lateral slope of the road is $\theta$, as shown in Fig. \ref{model}, and the acceleration due to gravity is denoted by $g$, the vehicle states are governed by the equations
\begin{gather}\label{dyn_2}
\begin{aligned}
m\dot{v}+ mu\dot\psi+mg\sin\theta = F_{yf}+F_{yr}\\
{I}\ddot{\psi} =l_fF_{yf}-l_rF_{yr}.
\end{aligned}
\end{gather}

\noindent A detailed derivation of these equations can be found in \cite{bhardwaj2019estimating}.

\subsection{Tire Model}

The rack force is primarily influenced by the aligning moments acting on the front tires of a vehicle because the movement of the steering rack is linked to the steering angle of the front tires. Therefore, to enable rack force estimation, the primary objective of a tire model is to estimate the aligning moments acting on the front tires. For the bicycle model, since the front tires are lumped into a single tire, the goal of the tire model is simply to estimate the aligning moment $M_{zf}$.

In order to obtain the aligning moment, it is first required to obtain tire slip angles and tire normal forces. The vehicle states obtained using Equations (\ref{dyn_1}) and (\ref{dyn_2}) can be used to determine the lateral slip angles $\alpha_f$ and $\alpha_r$ of the front and rear tires using the equations
\begin{gather}\label{alp}
\begin{aligned}
\alpha_f = \frac{v+l_f\dot{\psi}}{u}-\delta,\\
\alpha_r = \frac{v-l_r\dot{\psi}}{u}.
\end{aligned}
\end{gather}
The tire normal forces for the front tires $F_{zf}$ and for the rear tires $F_{zr}$ for driving on the road slope $\theta$ are given by the following equations
\begin{gather}\label{n_33}
\begin{aligned}
F_{zf} = \frac{mgl_r\cos\theta}{2(l_f+l_r)},\\ F_{zr} = \frac{mgl_f\cos\theta}{2(l_f+l_r)}. 
\end{aligned}
\end{gather}

The slip angles presented in Equation (\ref{alp}) and normal forces presented in Equation (\ref{n_33}) remain the same in the tire models which are discussed next.

\subsubsection{Linear Tire (LT) Model}\label{LT}
The lateral forces on the front and rear tires with respective cornering stiffness $C_{\alpha_f}$ and $C_{\alpha_r}$ are given by 
\begin{gather}\label{LT_fo}
F_{yf} = C_{\alpha_f}\alpha_f;\hspace{1em}
F_{yr} = C_{\alpha_r}\alpha_r.
\end{gather}

To express the lateral forces acting on both the front ($f$) and rear ($r$) tires using a single equation, let us rewrite Equation (\ref{LT_fo}) in the following form 
\begin{gather}\label{LT_f}
F_{yi} = C_{\alpha_i}\alpha_i,
\end{gather}
where $i\in \lbrace{f,r}\rbrace$.

\noindent The front tire pneumatic trail $t_p$ for the LT Model is given by the expression \cite{balachandran2014virtual}
\begin{gather*}
    t_p = t_{p0}\left(1-\sgn(\alpha_f) \frac{C_{\alpha_f}}{3\mu F_{zf}}\tan \alpha_f\right),
\end{gather*}
where $t_{p0}$ is the pneumatic trail at zero front slip angle.

The aligning moment $M_{zf}$ for the front tire is then given by
\begin{gather}\label{LT_m}
M_{zf} = -(t_p+t_m)F_{yf},
\end{gather}
where the front tire mechanical trail $t_m$ is a constant for a given vehicle. For a detailed description of the pneumatic trail $t_{p}$ and the mechanical trail $t_{m}$, the reader is referred to \cite{pacejka2005tire}.

Note that the tire lateral forces are directly proportional to the slip angles which is why this model is referred to as the ``linear tire" model. Such proportionality only applies within a range of slip angles, in particular for low values of slip angles, after which it no longer captures the variation of tire forces accurately \cite{pacejka2005tire}.

\subsubsection{Brush Tire (BT) Model}\label{BT}
 
Unlike the LT Model where the tire forces vary linearly with the slip angles, in the BT Model the tire forces are non-linear in the slip angles. Therefore, the BT Model can provide a better estimate of rack force over a larger range of slip angles. According to \cite{pacejka2005tire}, the lateral tire force for the front ($f$) and rear ($r$) tires is given by 

\begin{gather}\label{3dof_F}
\hspace{-0.7em}F_{yi} = \left\{
	\begin{array}{ll}
		\mu F_{zi}(3\theta_s\alpha_i - 3(\theta_s\alpha_i)^2 + (\theta_s \alpha_i)^3)  & \mbox{if } \alpha_i\leq \frac{1}{\theta_s} \\
		\mu F_{zi} & \mbox{if } \alpha_i\geq \frac{1}{\theta_s},
	\end{array}
\right.
\end{gather}
where again $i\in \lbrace{f,r}\rbrace$ and $\mu$ is the coefficient of friction between tire and road. The normal force $F_{zi}$ is given by Equation (\ref{n_33}), and $\theta_s$ is a tire parameter that, for tire tread stiffness of $c_p$ and the contact patch length of $2a$, is defined by 
\begin{align*}
\theta_s = \frac{2}{3}\frac{c_pa^2}{\mu F_{zi}}
\end{align*}
The front tire pneumatic trail $t_p$ for the BT Model is given by the expression \cite{pacejka2005tire}
\begin{gather}
t_p = \frac{1}{3}a\frac{1-3\abs{\theta_s\alpha_f}+3(\theta_s\alpha_f)^2-\abs{\theta_s\alpha_f}^3}{1-\abs{\theta_s\alpha_f}+\frac{1}{3}(\theta_s\alpha_f)^2}.
\end{gather}
For mechanical trail $t_m$, the tire aligning moment $M_{zf}$ can then be obtained by using the following expression which is the same as the one used in the LT Model
\begin{gather}\label{BT_m}
M_{zf} = -(t_p+t_m)F_{yf}.
\end{gather}

\subsubsection{Rigid Ring (RR) Model}\label{RR}

The LT and BT Model can estimate the tire forces and moments for low frequency road profile variations ($<$8Hz) such as road slopes. The Rigid Ring (RR) tire model is designed to estimate tire forces and moments for higher frequency road profile changes (such as 8Hz$-$80Hz) \cite{schmeitz2004semi, zegelaar1996plane, pacejka2005tire}. Moreover, like the BT Model, the RR Model can capture nonlinear dependence of tire forces on slip angles better than the LT Model. 

In the RR Model, the profile or the geometry of the road surface are first used to determine an ``effective road profile" using a tire enveloping model \cite{schmeitz2004semi}. The effective road profile is then used to compute the contact patch normal force $F_{cN}$. A detailed derivation of $F_{cN}$ can be found in \cite{bhardwaj2020rack, schmeitz2004semi}. The tire lateral forces $F_{yi}$ (where $i\in \lbrace{f,r}\rbrace$) can then be estimated using the expression \cite{schmeitz2004semi}
\begin{gather}\label{RR_f}
\begin{aligned}
F_{yi} = D_y \sin(C_y \arctan\lbrace B_y\alpha_{yi} - E_y(B_y\alpha_{yi}- \arctan(B_y\alpha_{yi}))\rbrace)+S_{Vy}.
\end{aligned}
\end{gather}
The front tire pneumatic trail $t_p$ for the RR Model is given by
\begin{gather*}
t_p = D_t \cos(C_t \arctan\lbrace B_t\alpha_{tf} - E_t(B_t\alpha_{tf}-\arctan(B_t\alpha_{tf}))\rbrace),
\end{gather*}
and the resulting aligning moment $M_{zf}$ acting on the front tire can be estimated using
\begin{gather}\label{RR_m}
    M_{zf} = -t_pF_{yf}+D_r\cos(\arctan{B_r\alpha_{rf}}),
\end{gather}
where the slip angles ($\alpha_{yi}$, $\alpha_{ti}$, and $\alpha_{ri}$) are given by
\begin{gather*}
\alpha_{yi} = S_{Hy}+ \tan{\alpha_i}, \hspace{2em} 
\alpha_{ti} = S_{Ht}+\tan{\alpha_i}, \hspace{2em} 
\alpha_{ri} = \tan{\alpha_i}.
\end{gather*}

The coefficients $B_y$, $B_r$, $B_t$, $C_y$, $C_t$, $D_y$, $D_r$, $D_t$, $E_y$, $E_t$, $S_{Hy}$, and $S_{Ht}$ are either constants or are functions of slip angles ($\alpha_{yi}$, $\alpha_{ti}$, and $\alpha_{ri}$) and contact patch normal forces $F_{cN}$, and tire normal forces $F_{zi}$ \cite{bhardwaj2020rack,schmeitz2004semi}.

\subsection{Rack Force Estimation}

The aligning moment $M_{zf}$ obtained using each model was used to estimate the resultant road feedback or the rack force $RF$ using the expression
\begin{gather}\label{fr}
RF = i_pM_{zf},
\end{gather}
where the constant ratio $i_p$ defines the tire moment to rack force transmission ratio for a given vehicle.

\subsection{Model Assembly}

Referring again to Fig. \ref{Fig:overall3}, the vehicle states (lateral speed $v$ and yaw rate $\dot{\psi}$) are produced by the vehicle model represented by Equation (\ref{dyn_1}) for flat roads and Equation (\ref{dyn_2}) for sloped roads. Independently, the vehicle's mass and dimensions and the road slope are used to compute normal tire forces using Equation (\ref{n_33}). In the tire model, the vehicle states are used to find the tire slip angle for the front and rear tires using Equation (\ref{alp}). Tire slip angles and normal forces are then used to find tire forces and aligning moments using Equations (\ref{LT_f}) and (\ref{LT_m}) for the LT Model, Equations (\ref{3dof_F}) and (\ref{BT_m}) for the BT Model, and Equations (\ref{RR_f}) and (\ref{RR_m}) for the RR Model. The tire forces are fed back into the vehicle model in Equations (\ref{dyn_1}) and (\ref{dyn_2}) to generate the vehicle states for the next time instant. During this process, the rack force for each time instant is obtained through Equation (\ref{fr}) using the aligning moment estimated in Equation (\ref{LT_m}) for the LT Model, in Equation (\ref{BT_m}) for the BT Model, and in Equation (\ref{RR_m}) for the RR Model. 

\subsection{Targeted Compensation using Rack Force Components}

VTM-based estimators can be utilized to determine the rack force due to steering angle independent of the rack force due to road profile. The components can then be used to compensate for the individual effects of steering angle and road profile on the steering feel. However, there are some preliminary requirements for performing such a targeted compensation. In this subsection, we attempt to outline these requirements. 

Consider rack force $RF$ obtained in Equation (\ref{fr}). $RF$ is clearly a nonlinear function of steering angle and road profile, and can be expressed using a generic function $f$ as 
\begin{gather}\label{slo_l2}
RF = f(\delta,\theta).
\end{gather}
Now let us denote rack force due to steering angle by $RF_{Steering}$ and rack force due to road profile by $RF_{Road}$. Using Equation (\ref{slo_l2}), $RF_{Steering}$ and $RF_{Road}$ can be obtained by using one input at a time in function $f$ as follows
\begin{gather}\label{slo_l3}
    RF_{Steering} = f(\delta,0), \hspace{1em} RF_{Road} = f(0,\theta).
\end{gather}

Since rack force is nonlinear, $RF_{Steering}$ and $RF_{Road}$ may not be the only components of rack force. Rack force may consist of additional nonlinear components arising from the interaction of steering angle and road profile inputs. Let us combine the additional components of rack force into a single variable $\Delta RF$ which we call the residual rack force. The rack force $RF$ can then be decomposed into three components

\begin{gather}\label{slo_l4}
    RF = RF_{Steering} + RF_{Road} + \Delta RF
\end{gather}

Clearly, to properly perform targeted compensation it is important to verify whether the residual rack force $\Delta RF$ is small in comparison to $RF_{Steering}$ and $RF_{Road}$. If $RF_{Steering}$ and $RF_{Road}$ are not the primary components of rack force, compensating only for the effect of steering angle and road profile may not be sufficient. The unknown and uncompensated residual rack force may result in incorrect compensation which in turn may result in a substandard or undesirable steering feel. In Section \ref{sec:results}, we compare the sum of independent estimates $RF_{Steering}$ and $RF_{Road}$ with rack force $RF$ to check whether rack force is primarily composed of only $RF_{Steering}$ and $RF_{Road}$. 

To perform targeted compensation, it is also important to verify whether the estimates of rack force due to steering angle $RF_{Steering}$ and rack force due to road profile $RF_{Road}$ obtained using the estimator accurately represent the contributions of steering angle and road profile to rack force. 
Unfortunately, unlike the total rack force $RF$, the component-wise estimates of rack force $RF_{Steering}$ and $RF_{Road}$ cannot be validated using the force measurements available from strain gauges mounted on the steering rack. However, estimates produced by higher DOF VTM-based rack force estimators, such as those available in commercial vehicle dynamics packages, can still serve as a reference to compare the estimates produced by low DOF VTM-based rack force estimators. In the next section we discuss how the component-wise estimates produced by one of our estimators were compared with the estimates produced by a higher DOF rack force estimator available in CarSim.

\section{Methods}\label{sec:methods}

We now describe the experimental setup and the simulation setup used to test the performance of the three rack force estimators developed in this paper. We first describe the three driving experiments and the hardware setup that were used to determine and compare the real-time estimation accuracies of the rack force estimators. After that, we explain the simulation setup that was used to verify whether the component-wise estimates of rack force can be used to perform targeted compensation.

\subsection{Experimental setup}\label{EX_ST}
Driving experiments were performed at test tracks with known road profile variations. The test tracks were located at Ford's Dearborn Development Center (formerly Dearborn Proving Grounds) in Dearborn, Michigan. The results from the following experiments are described in this paper:

\begin{enumerate}
    \item \textit{Experiment 1: Driving on a road with varying lateral slope}\\
    This driving experiment was performed on a crowned road with $11^{\circ}$ slope on the two sides of the road crown. The vehicle was driven from one side to the other side with a speed of about 20 km/h.
    \item \textit{Experiment 2: Aggressive slalom driving on a road with constant lateral slope}\\ 
    This experiment was performed on a road with constant lateral slope of about $11^{\circ}$. The steering angle was varied between approximately $-60^{\circ}$ and $60^{\circ}$ to perform a slalom maneuver with a speed of about 15 km/h.
    \item \textit{Experiment 3: Slalom driving on a road with cleats of varying heights}\\
    This experiment was performed on a road with thirteen metal cleats of known dimensions: the first four cleats were 1 cm tall, the next five cleats were 2 cm tall, and the remaining cleats were 3 cm tall. All cleats were 4 cm long and were oriented transverse to the road. Driving speed was maintained at about 30 km/h. The steering angle was varied between approximately $-30^{\circ}$ and $30^{\circ}$ so that the vehicle impacted the cleats at an angle (roughly equal to the steering angle). Such a maneuver was performed to test the performance of the estimators for driving over arbitrary high frequency unevenness, such as oblique cleats, on the road. Moreover, hitting the cleats straight with zero impact angle did not significantly influence the tire aligning moment and therefore did not induce much rack force.
\end{enumerate}

\begin{figure}[h!]
\centering
\includegraphics[width=0.8\textwidth]{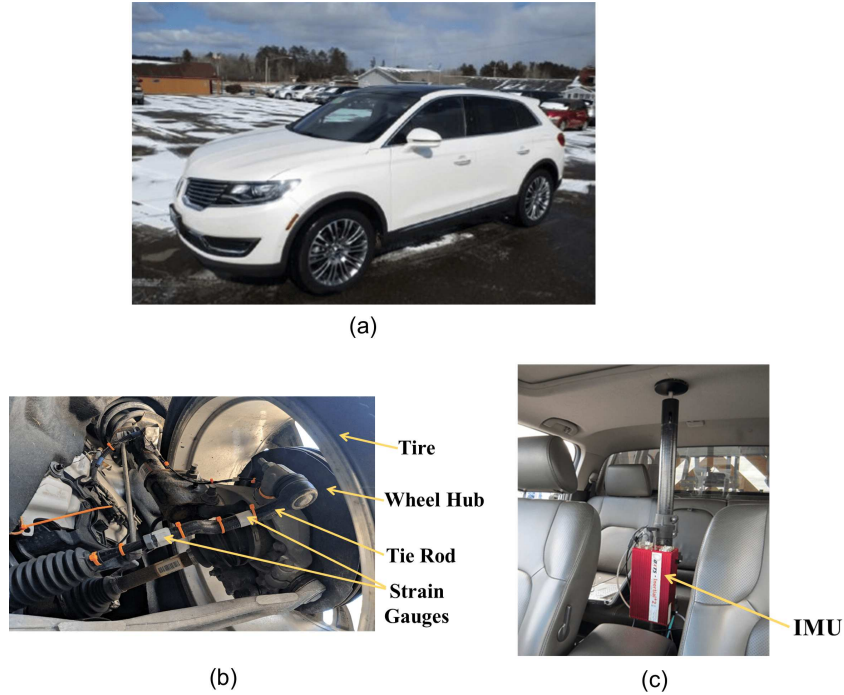}
\caption{Experimental setup. (a) Lincoln MKX test vehicle. (b) Tie rod instrumented with strain gauges to measure the steering rack force. (c) Inertial measurement unit (IMU) mounted inside the car used to measure the road slopes during driving.}\label{test_set}
\end{figure}
The experiments were performed using the Lincoln MKX vehicle equipped with Pirelli (Scorpion Verde AS) tires shown in Fig. \ref{test_set} (a). Tire model specific parameters used in the models were taken from \cite{pacejka2005tire} and \cite{schmeitz2004semi}. Other vehicle specific parameters can be found in \cite{bhardwaj2019estimating}. To evaluate the estimation performances of the models, the rack force estimates produced by the three models were compared to measurements from strain gauges installed on the tie rods of the test vehicle (shown in Fig. \ref{test_set} (b)). The steering angle and the vehicle speed were measured using steering angle and tire speed sensors, respectively. The road slopes (longitudinal and lateral) were obtained using the pitch and roll measurements obtained from a high fidelity IMU (OXTS RT3003 v2) installed in the vehicle (shown in Fig. \ref{test_set} (c)) that transmitted signals at 100 Hz. The pitch and roll measurements were assumed to be roughly equal to the longitudinal and lateral slopes of the road. The cleat dimensions (height, width and length) were physically measured on the track where the tests were performed. During the driving tests, a rapid control prototyping platform (dSPACE MicroAutoBox) was used to link sensed steering angle, road profile, and vehicle speed signals with an online simulation of the three rack force estimators (integrated in real-time Simulink), using CAN-bus communications at 250 Hz. 

\subsection{Simulation Setup}\label{CS_set}

The BT Model was used to produce the component-wise estimates of rack force: rack force due to steering angle $RF_{Steering}$ and rack force due to road profile $RF_{Road}$. 
Unlike the estimates of total rack force, the component-wise estimates of rack force could not be measured using sensors available in the vehicle. Therefore, a higher DOF VTM-based estimator available in CarSim was chosen as a reference to validate the component-wise estimates of rack force produced by the BT Model. 
The VTM-based estimator in CarSim had a four-wheel vehicle model that had 15 mechanical degrees of freedom (DOF) in comparison to the bicycle model with only two DOF used in the BT Model. The math model for the 15 DOF model vehicle in CarSim had over 250 state variables \cite{carsim2017math}. For tire models, CarSim provided various options for tire models that had higher complexity than the Brush Tire model used in the BT Model. We used the semi-empirical tire model called ``Internal Table Model with Simple Camber" which used combined slip theory \cite{bakker1989new} and similarity method \cite{pacejka1991shear} to compute tire forces and moments \cite{sayers2018tire}. 

To perform the simulation experiment, we recreated Experiment 1 described in Section \ref{EX_ST} in the CarSim environment. We fed the recorded steering angle from Experiment 1 into the CarSim Simulink Model and re-created the same road profile in the CarSim driving environment that was traversed while performing Experiment 1. Moreover, we performed the simulation on an SUV vehicle with the vehicle dimensions, mass, yaw inertia, and tire size similar to the Lincoln MKX vehicle on which the physical test was performed (as shown in Table \ref{table0}). 
\begin{table}[h!]
 \setlength\extrarowheight{2pt}
\centering
\caption{Comparison of parameters between the test vehicle and simulated vehicle}
 \begin{tabular}{||c c c||} 
 \hline
 Parameters/Vehicle & Test Vehicle & Simulated Vehicle \\ [0.5ex] \hline\hline
 Mass (kg) & 1972 & 2257\\
 Yaw Inertia (kg-m\textsuperscript{2}) & 3600 & 3525 \\
 Track Width (m) & 1.64 & 1.69\\
 Wheel Base (m) & 2.88 & 2.95\\
 Tire Size & 265/40 R21 & 255/45 R19 \\
  \hline
\end{tabular}
\label{table0}
\end{table}

\begin{figure}[h!]
    \centering
    \includegraphics[width=\textwidth]{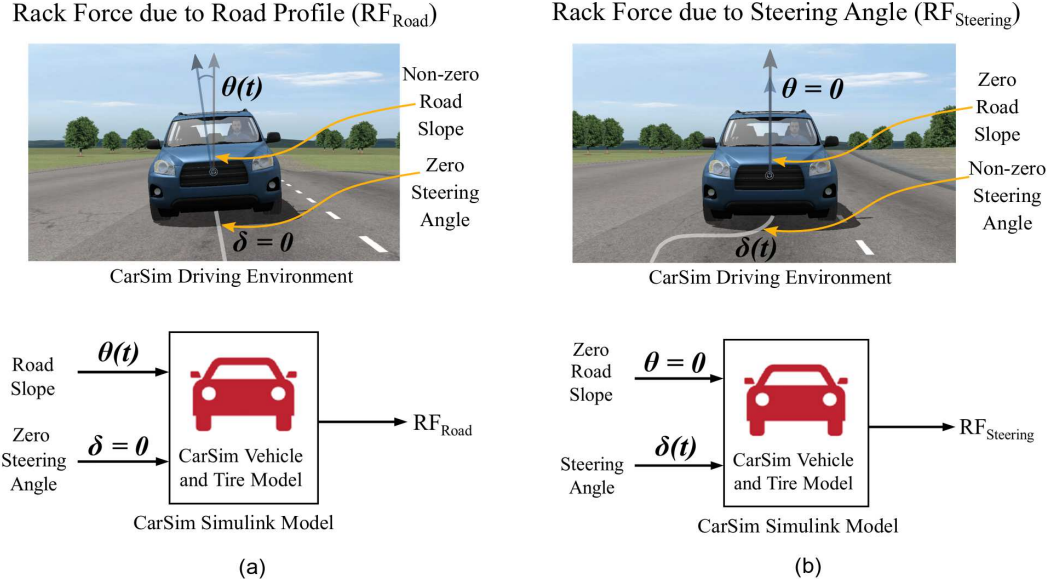}
    \caption{CarSim setup for component-wise rack force estimation. (a) $RF_{Road}$ was estimated by setting the steering angle to zero in the CarSim Simulink model and by creating a sloped road profile in the driving environment. (b) $RF_{Steering}$ was estimated by making the steering angle non-zero in the CarSim Simulink model and by creating a flat road profile in the driving environment.}
    \label{css_set}
\end{figure}

The CarSim simulation setup is summarized in Fig. \ref{css_set}. The first simulation experiment was used to estimate $RF_{Road}$ where the steering angle was set to zero but the driving was simulated on a sloped road (shown in Fig. \ref{css_set} (a)). The second simulation experiment was used to estimate $RF_{Steering}$ where the steering angle was non-zero but the driving was simulated on a flat road (shown in Fig. \ref{css_set} (b)). The vehicle speed remained the same in both simulation experiments. The component-wise estimates obtained using CarSim were then compared with the estimates obtained using the BT model. Finally, the sum of component-wise estimates obtained using the BT Model was compared with the total rack force estimate to determine the contribution of residual rack force to the total rack force.

\subsection{Performance Analysis}

To quantify the performance of a given estimator, we used the normalized mean absolute error (NMAE) between the rack force estimate produced by the estimator, $RF$, and the rack force estimated using a reference estimator, $RF_{ref}$. NMAE was expressed as a percentage and was obtained using the following equation
\begin{gather}\label{perfmet}
    \text{NMAE(\%)} = \frac{\text{mean}(\vert RF_{ref}-RF\vert)}{\text{max}(RF_{ref})-\text{min}(RF_{ref})}\times100
\end{gather}
For the driving experiments the sensors mounted in the vehicle served as the reference, whereas for simulation experiments the rack force estimates produced using CarSim served as the reference. We also compared the rack force estimation performance of our estimators to each other to find out which estimator had the highest relative accuracy. 

\section{Results and Discussion}\label{sec:results}

\subsection{Effect of Model Complexity on Rack Force Estimation}

The differences between the three VTM-based estimators were readily apparent in the comparison of estimation errors between the estimators as shown in Table \ref{table:1}. 
\begin{table}[h!]
 \setlength\extrarowheight{2pt}
\centering
\caption{Normalized mean absolute estimation errors (\%) for the three experiments}
 \begin{tabular}{||c c c c||} 
 \hline
 Estimator/Experiment & Experiment 1 & Experiment 2 & Experiment 3 \\ [0.5ex] 
  \hline\hline
 LT (Linear Tire) & 3.75\% & 10.11\% & 9.17\% \\
 BT (Brush Tire) & 3.83\% & 7.88\% & 8.66\% \\
 RR (Rigid Ring) & 3.38\% & 7.51\% & 5.35\% \\
  \hline
\end{tabular}
\label{table:1}
\end{table}

\begin{figure}[h!]
	\centering
	\begin{minipage}[b]{\minipagetextwidth\textwidth}
		\includegraphics[width=\textwidth]{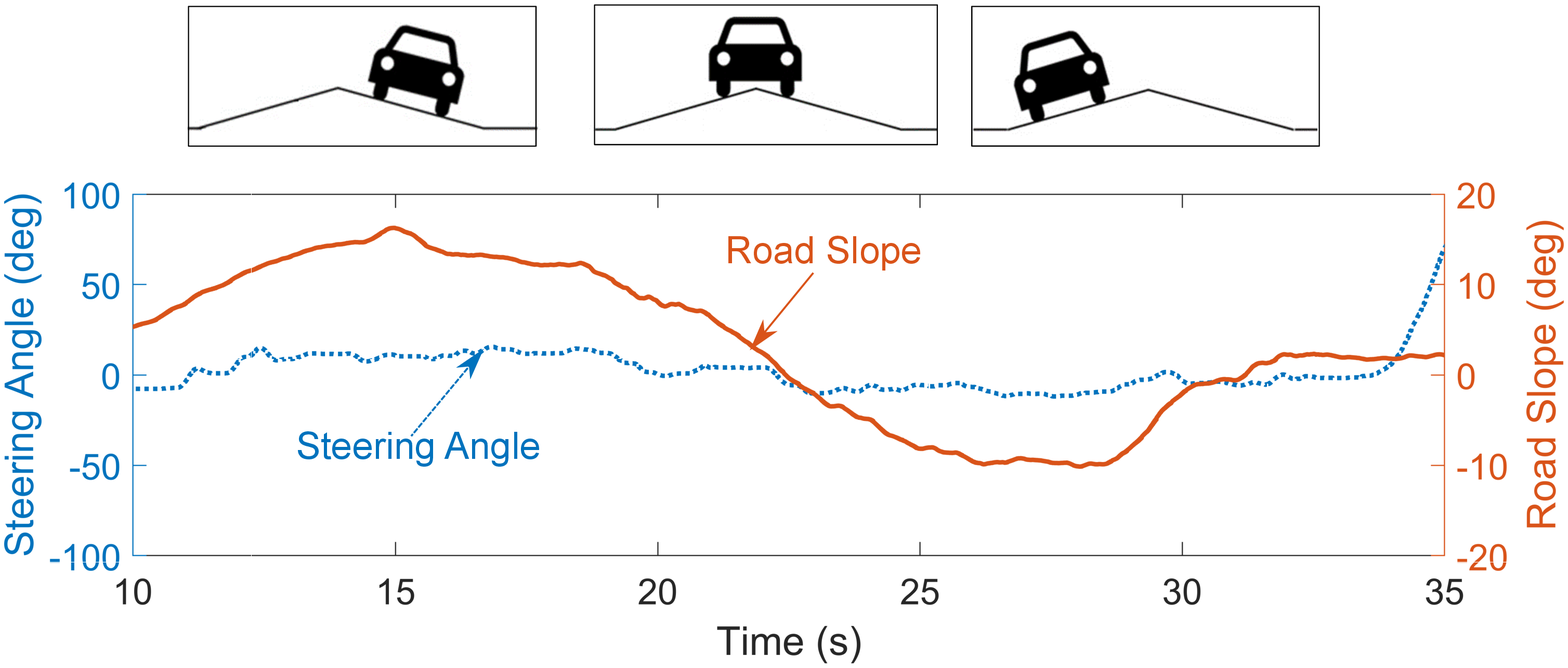}
 \subcaption{}
    \label{1a}
\end{minipage}
\begin{minipage}[b]{\minipagetextwidth\textwidth}
		\includegraphics[width=\textwidth]{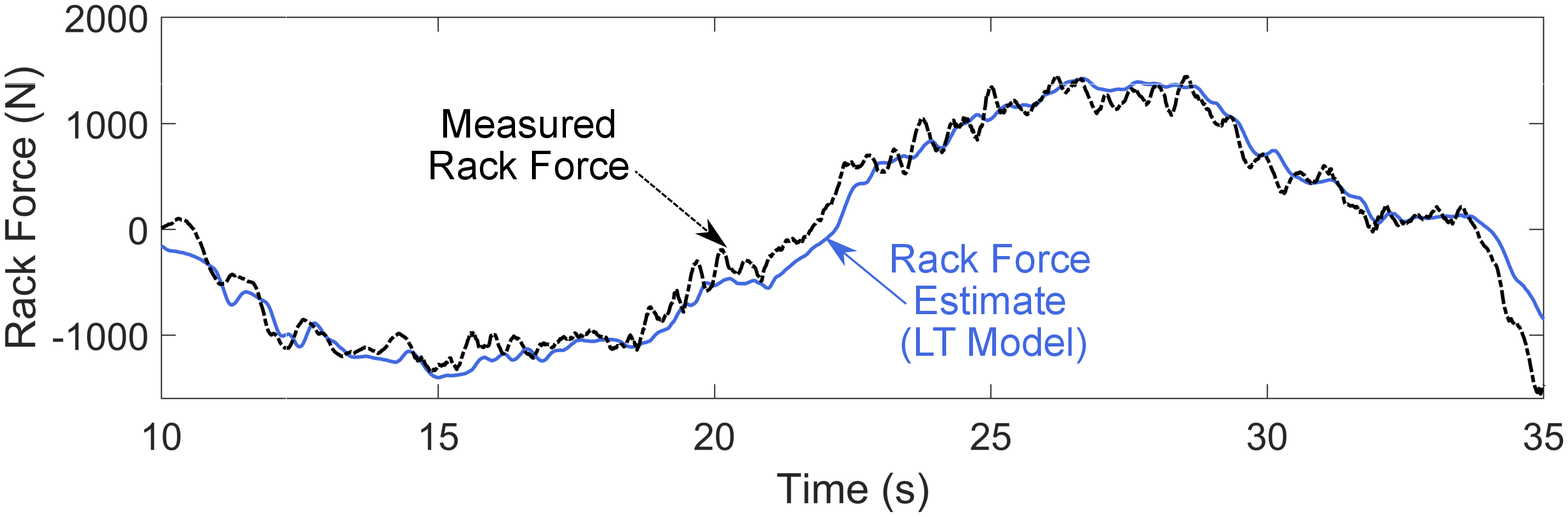}
 \subcaption{}
    \label{1b}
\end{minipage}
\begin{minipage}[b]{\minipagetextwidth\textwidth}
		\includegraphics[width=\textwidth]{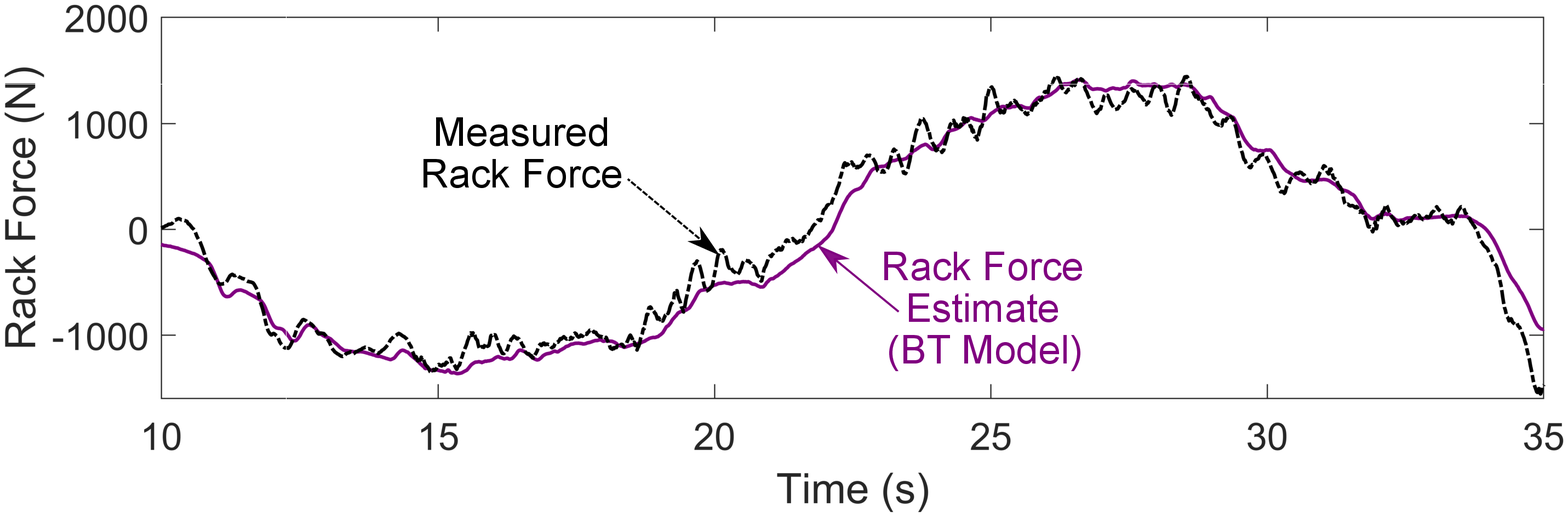}
\subcaption{}
    \label{1c}
\end{minipage}
\begin{minipage}[b]{\minipagetextwidth\textwidth}
		\includegraphics[width=\textwidth]{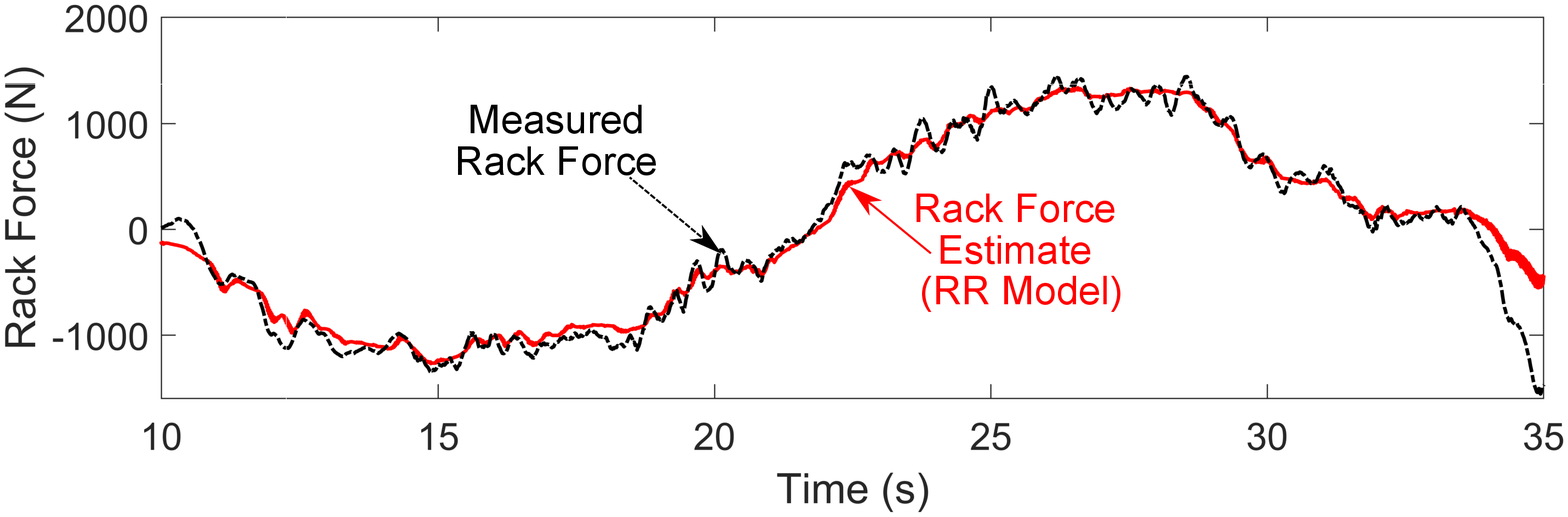}
\subcaption{}
    \label{1d}
\end{minipage}
	\caption{Experiment 1 performed on a crowned road. Vehicle speed was maintained at about 20 km/h. (a) Road profile variation (schematic) and input steering angle and road slope (graph). (b) Rack Force estimated using the LT Model and measured using sensor. (c) Rack Force estimated using the BT Model and measured using sensor. (d) Rack Force estimated using the RR Model and measured using sensor.} 
		\label{exp1}
\end{figure}

While driving on the road with large slope variation in Experiment 1, no differences were seen between the estimation performances.
Despite the significant distinction between the tire models used in our estimators, all estimators seemed to agree well with the sensor measurements (Fig. \ref{exp1}). The estimation errors were low for the three estimators and were only marginally different between the estimators (see Table \ref{table:1}). The results indicated that both the linear and nonlinear tire models were equally capable of estimating the rack force irrespective of the magnitude and variation of road slope.

\begin{figure}[h!]
	\centering
	\begin{minipage}[b]{\minipagetextwidth\textwidth}
		\includegraphics[width=\textwidth]{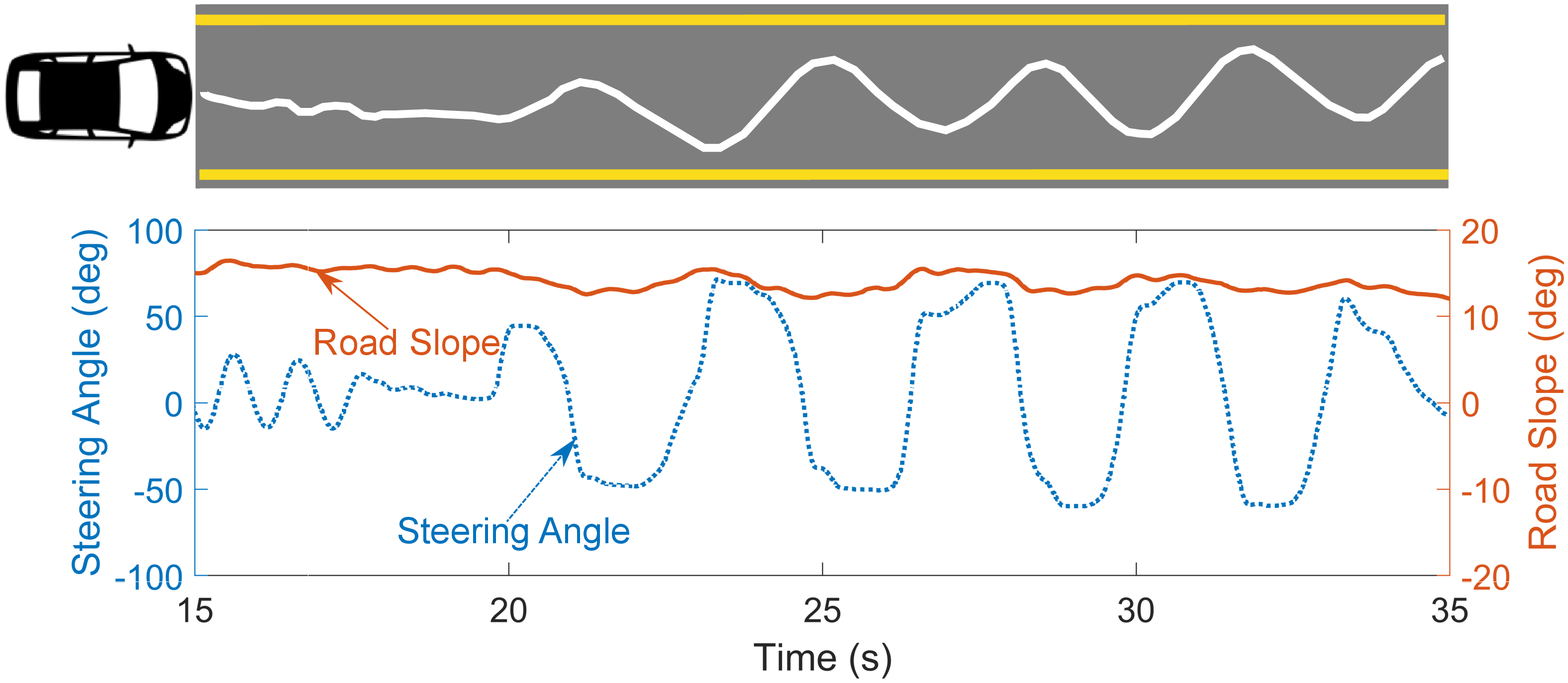}
 \subcaption{}
    \label{2a}
\end{minipage}
\begin{minipage}[b]{\minipagetextwidth\textwidth}
		\includegraphics[width=\textwidth]{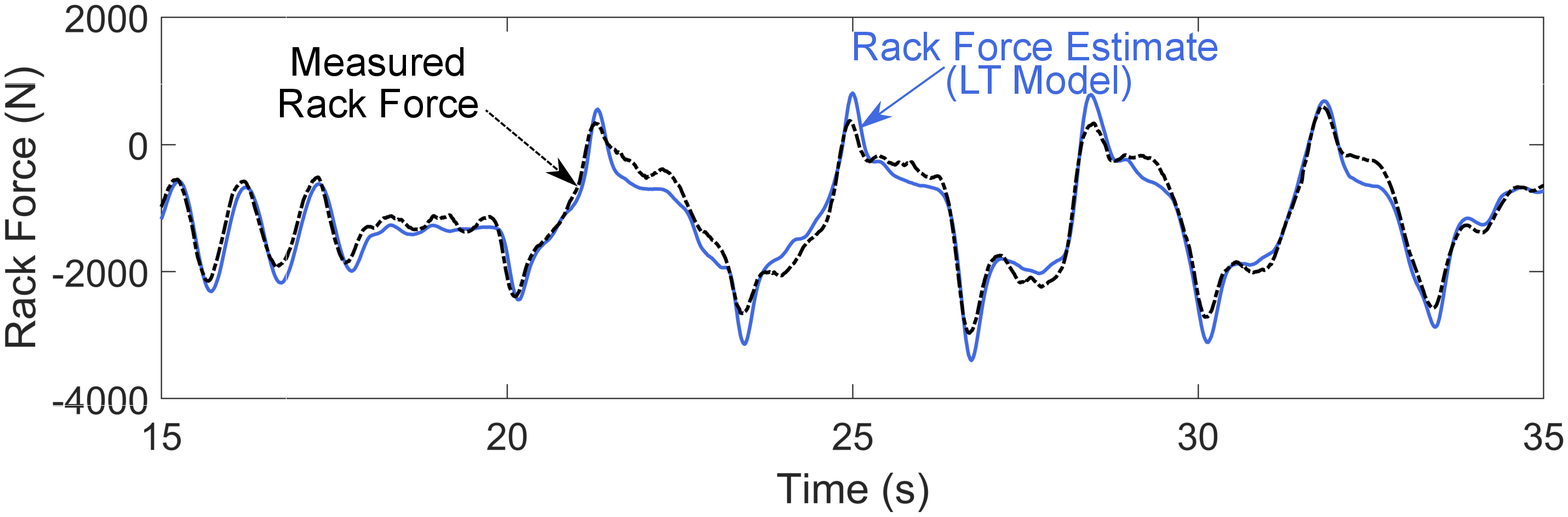}
 \subcaption{}
    \label{2b}
\end{minipage}
\begin{minipage}[b]{\minipagetextwidth\textwidth}
		\includegraphics[width=\textwidth]{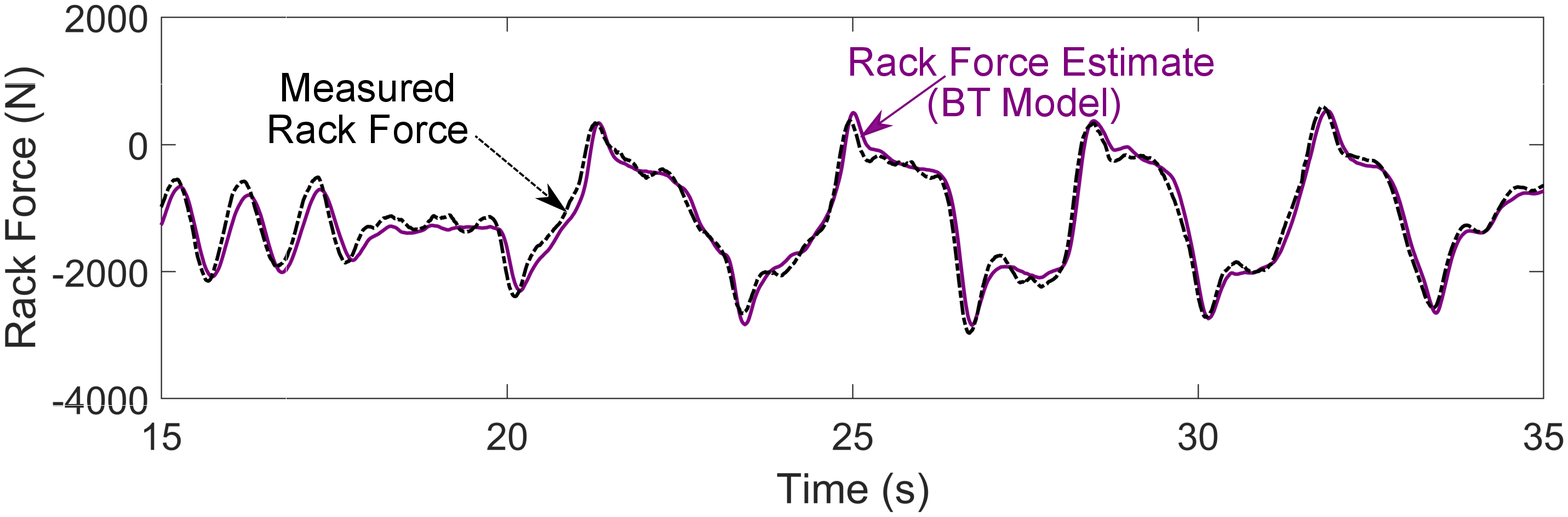}
\subcaption{}
    \label{2c}
\end{minipage}
\begin{minipage}[b]{\minipagetextwidth\textwidth}
		\includegraphics[width=\textwidth]{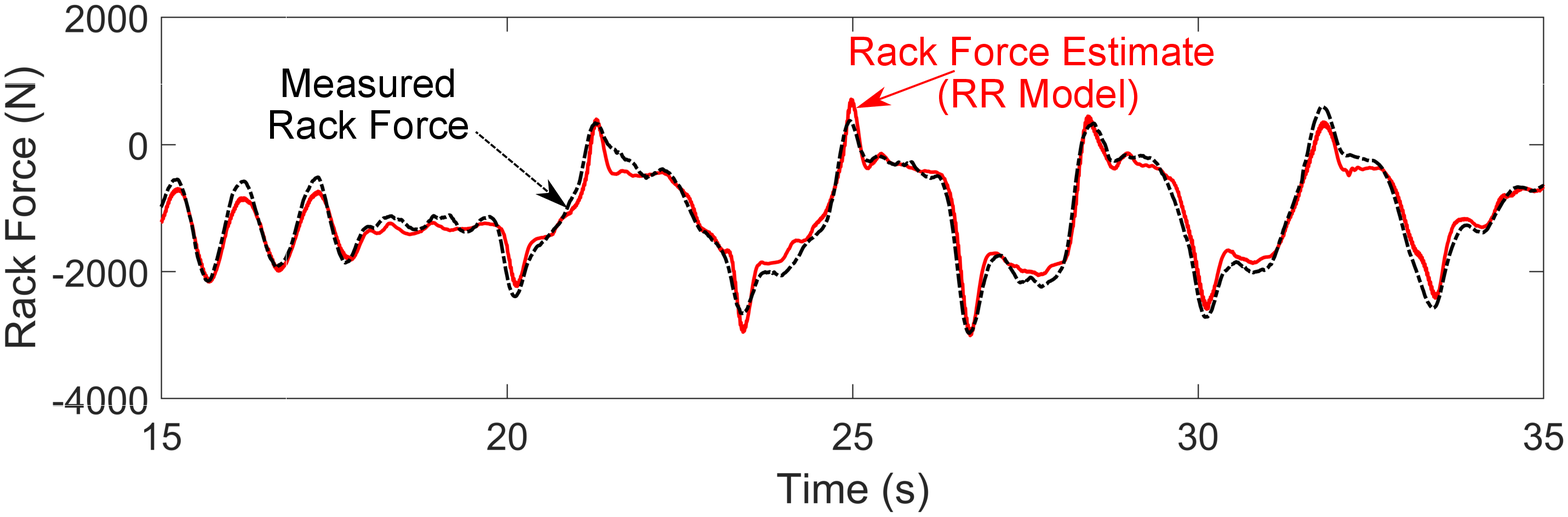}
\subcaption{}
    \label{2d}
\end{minipage}
	\caption{Experiment 2 performed with a slalom maneuver on a road with constant lateral slope. Vehicle speed maintained at about 15 km/h. (a) Vehicle maneuver (schematic) and input steering angle and road slope (graph). (b) Rack Force estimated using the LT Model and measured using sensor. (c) Rack Force estimated using the BT Model and measured using sensor. (d) Rack Force estimated using the RR Model and measured using sensor.} 
		\label{exp2}
\end{figure}

Driving with an aggressive slalom maneuver in Experiment 2, on the other hand, revealed some differences between the estimator using the linear tire model and the estimator using the nonlinear tire model. Estimators with nonlinear tire models, namely, the BT Model and the RR Model, matched the sensor measurements better than the LT Model (Fig. \ref{exp2}). For the LT Model, the estimation performance broke down at large steering angles and large steering angle rates (see Fig. \ref{2b}) while both the BT Model and the RR Model seemed to match the measurements well throughout the experiment (Fig. \ref{2c} and Fig. \ref{2d}). The normalized mean absolute estimation error for the RR Model and the estimation error for the BT Model were similar to each other and were both lower than the estimation error for the LT Model (see Table \ref{table:1}). 

\begin{figure}[h!]
	\centering
	\begin{minipage}[b]{\minipagetextwidth\textwidth}
		\includegraphics[width=\textwidth]{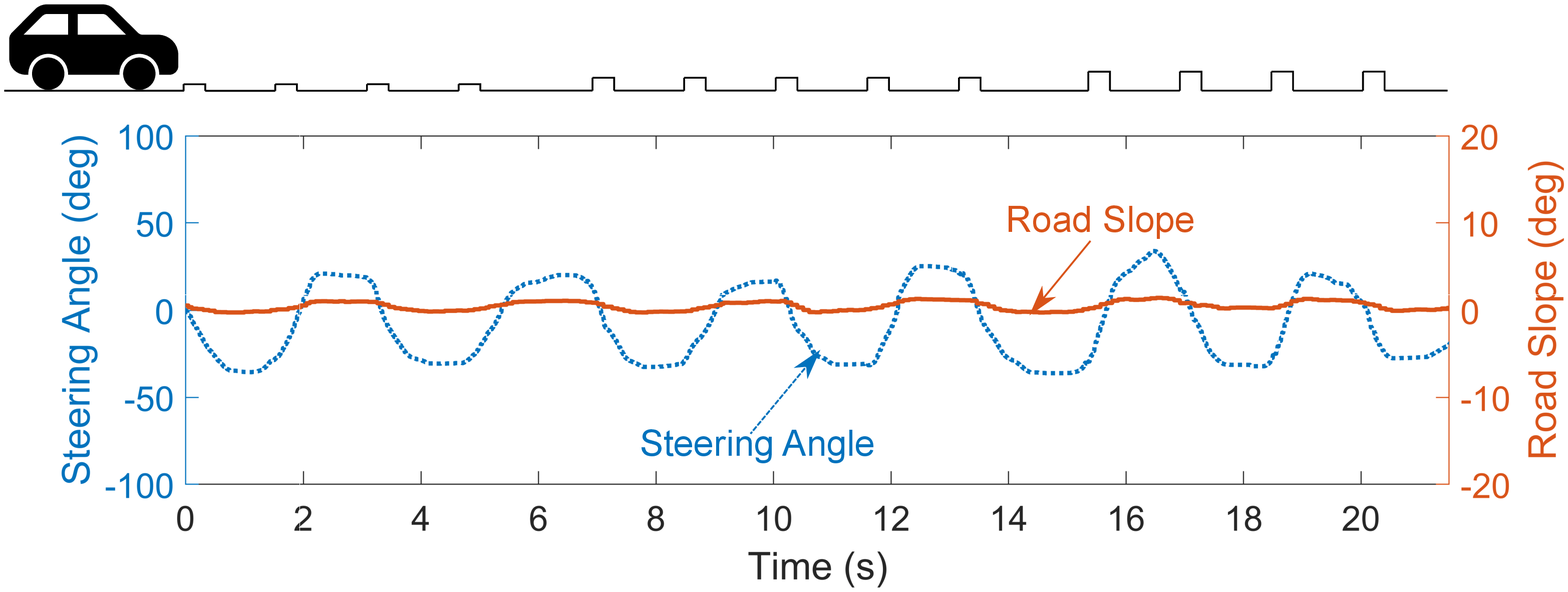}
 \subcaption{}
    \label{3a}
\end{minipage}
\begin{minipage}[b]{\minipagetextwidth\textwidth}
		\includegraphics[width=\textwidth]{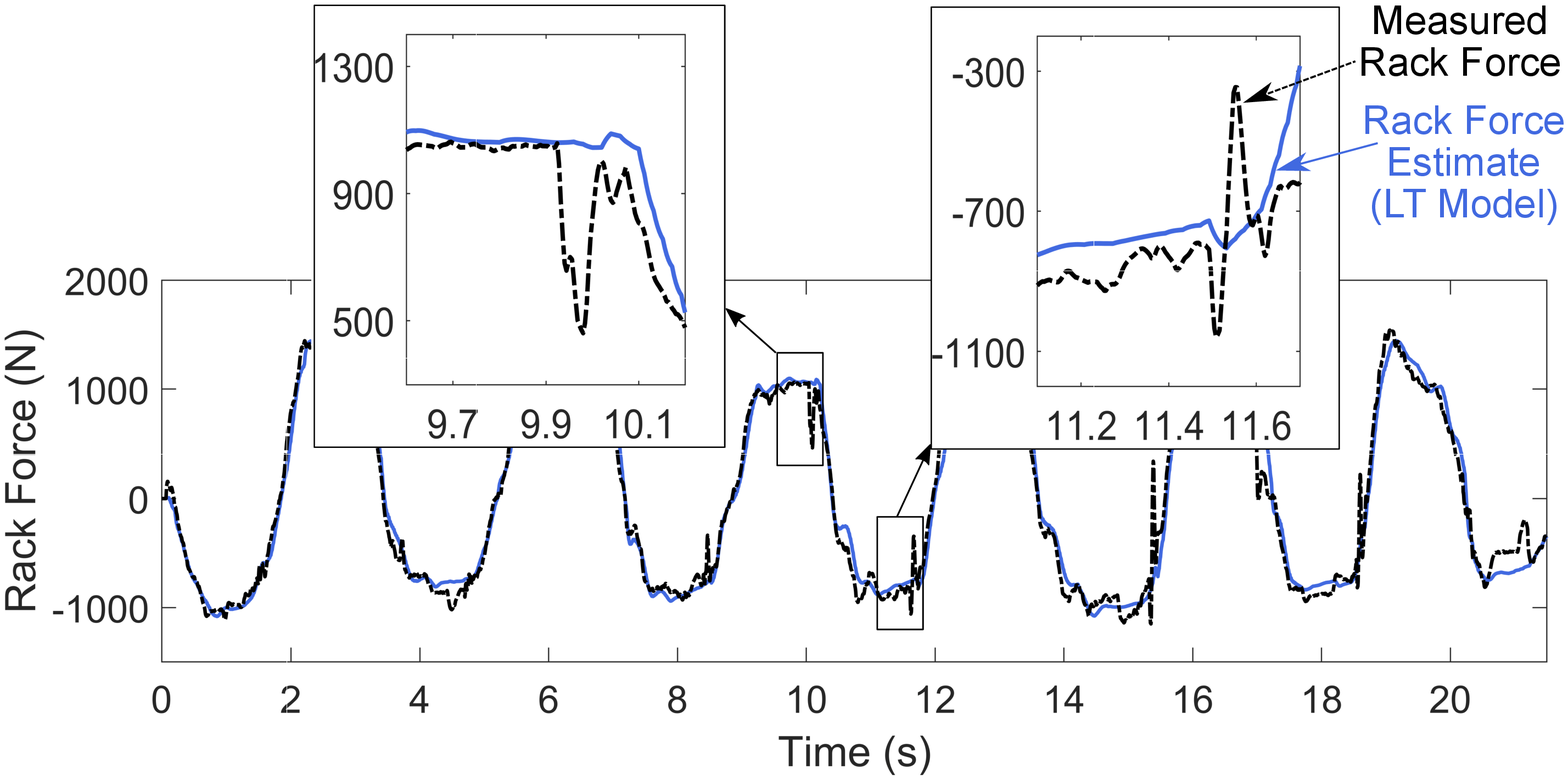}
 \subcaption{}
    \label{3b}
\end{minipage}
\begin{minipage}[b]{\minipagetextwidth\textwidth}
		\includegraphics[width=\textwidth]{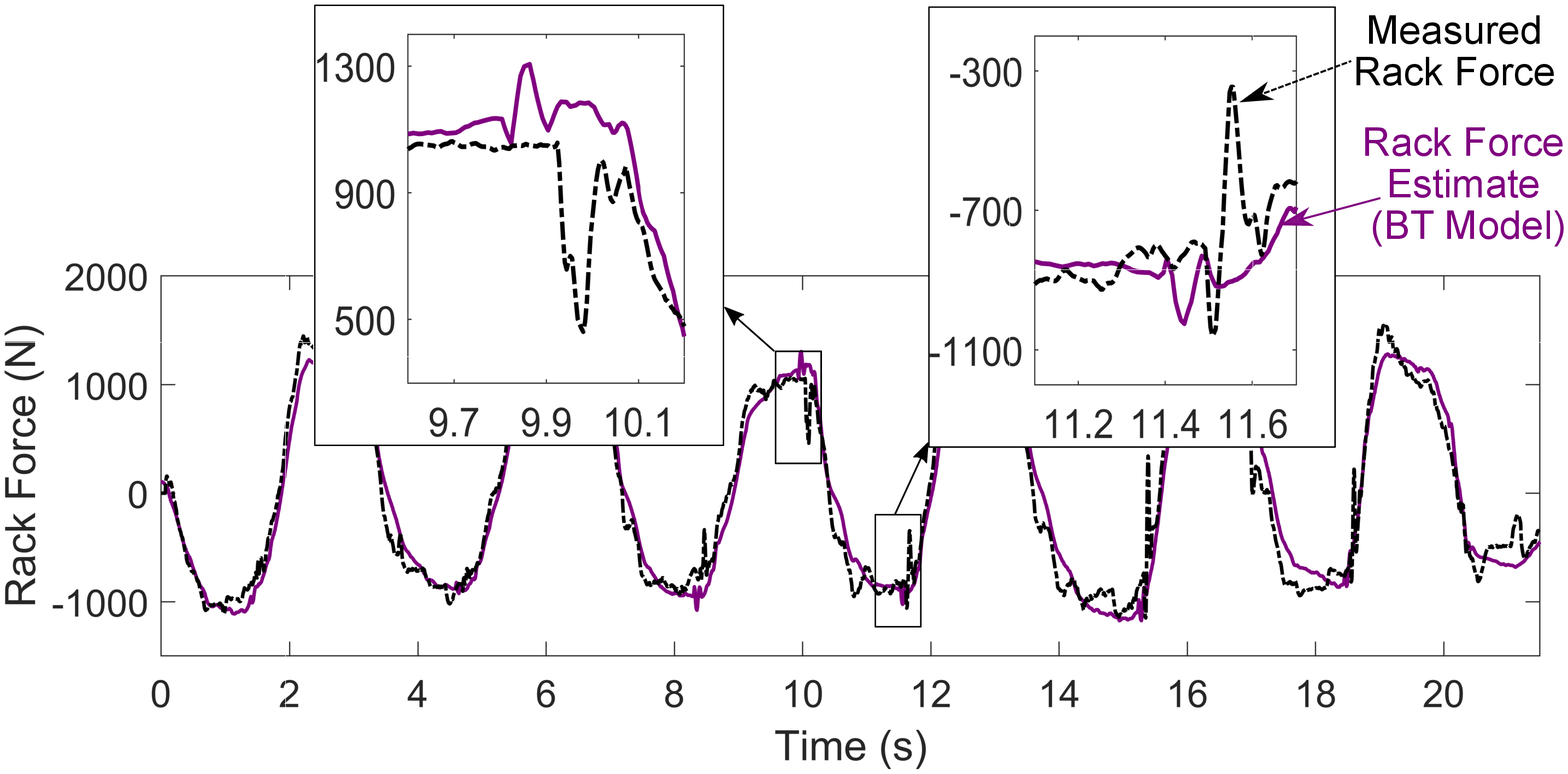}
\subcaption{}
    \label{3c}
\end{minipage}
\begin{minipage}[b]{\minipagetextwidth\textwidth}
		\includegraphics[width=\textwidth]{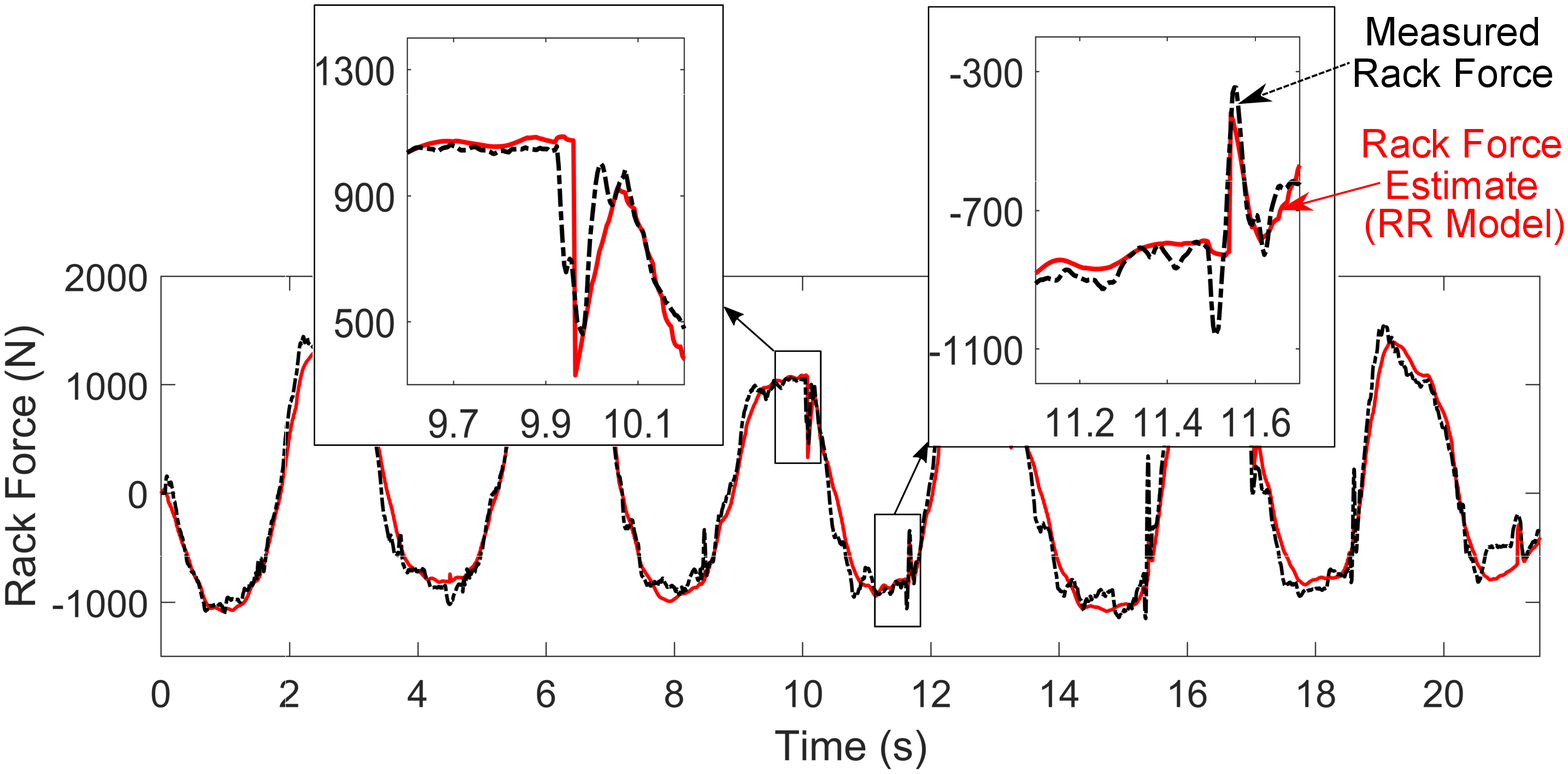}
\subcaption{}
    \label{3d}
\end{minipage}
	\caption{Experiment 3 performed with a slalom maneuver on a road with cleats of varying heights. Vehicle speed maintained at about 30 km/h. (a) Road profile variation (schematic) and input steering angle and road slope (graph). (b) Rack Force estimated using the LT Model and measured using sensor. (c) Rack Force estimated using the BT Model and measured using sensor. (d) Rack Force estimated using the RR Model and measured using sensor.} 
		\label{exp3}
\end{figure}

The estimation error for the RR Model was the lowest for driving on the road with cleats in Experiment 3. While driving on the flat part of the road, the performance of all estimators seemed similar (Fig. \ref{exp3}). However, only the RR Model captured the rack force well when the vehicle drove on cleats as demonstrated by the insets on the plots in Fig. \ref{exp3}. Furthermore, as shown in Table \ref{table:1}, the mean absolute estimation error for the RR Model was lower than both the BT Model and the LT Model. The BT Model and the LT Model, on the other hand, seemed to exhibit similar estimation performance. 

Clearly, our results reflect the differences in the complexity of tire models used in our estimators. For example, for Experiment 2, the RR Model and the BT Model perform better than the LT Model because the nonlinear tire models used in the BT Model and the RR Model are better at capturing the tire forces and aligning moments for higher steering angles and slip angles as compared to the linear tire model used in the LT Model. Likewise, in Experiment 3, the RR Model seems to have better performance than both the BT and the LT Model because only the RR Model accounts for high frequency road profile variations (such as cleats) in the estimation of tire forces and moments. And clearly this capability results in a nontrivial difference between the estimation errors of our models. On the other hand, all the estimators seem to exhibit satisfactory performance for non-aggressive steering maneuvers regardless of the magnitude and variation of the road slopes as demonstrated by Experiment 1. 

In other words, we found that in terms of driving on roads with low frequency profile variations ($<$8Hz) with non-aggressive steering maneuvers, all three estimators we developed seem to work sufficiently well. For driving with slalom steering maneuvers on low frequency road profile variations, however, both the BT Model and the RR Model outperform the LT Model. The BT Model appears to be a better choice for driving on low frequency road profile variations as it is computationally less intensive than the RR Model and produces rack force estimates of accuracy similar to the RR Model. However, for high frequency road profile variation such as produced while driving on road cleats, the RR Model appears to be a better choice for rack force estimation as it supports better estimation performance than the other models.

The VTM-based estimators developed in this paper are capable of estimating rack force for driving on different types of road profiles and can therefore be utilized to develop and improve various driver assist controllers. The estimators can be used in existing driver assist controllers that temporarily deactivate their functions when the vehicle transitions from a flat road to an uneven road (see, for example, controllers designed in \cite{dornhege2017steering, pick, raad2014pull, shah2011pull}). The estimators can also be used in virtual prototyping to analyse ride-comfort and durability of a vehicle and in simulating road feedback in hardware-in-the-loop simulators and simulation experiments \cite{bosch2002tyre, segawa2006preliminary, izadi2020impedance}. The RR Model, in particular, can also be used in semi and fully autonomous vehicles equipped with advanced road preview technology. Using the road profile inputs available from road preview sensors, the RR Model can enable pre-emptive estimation of rack force and modulation of steering torque feedback while driving on roads with slopes, cleats, or potholes \cite{gordon2015automated}. 

\subsection{Estimation of Rack Force Components}\label{Res_DC}

\begin{figure}[h!]
	\centering
\begin{minipage}[b]{\minipagetextwidth\textwidth}
		\includegraphics[width=\textwidth]{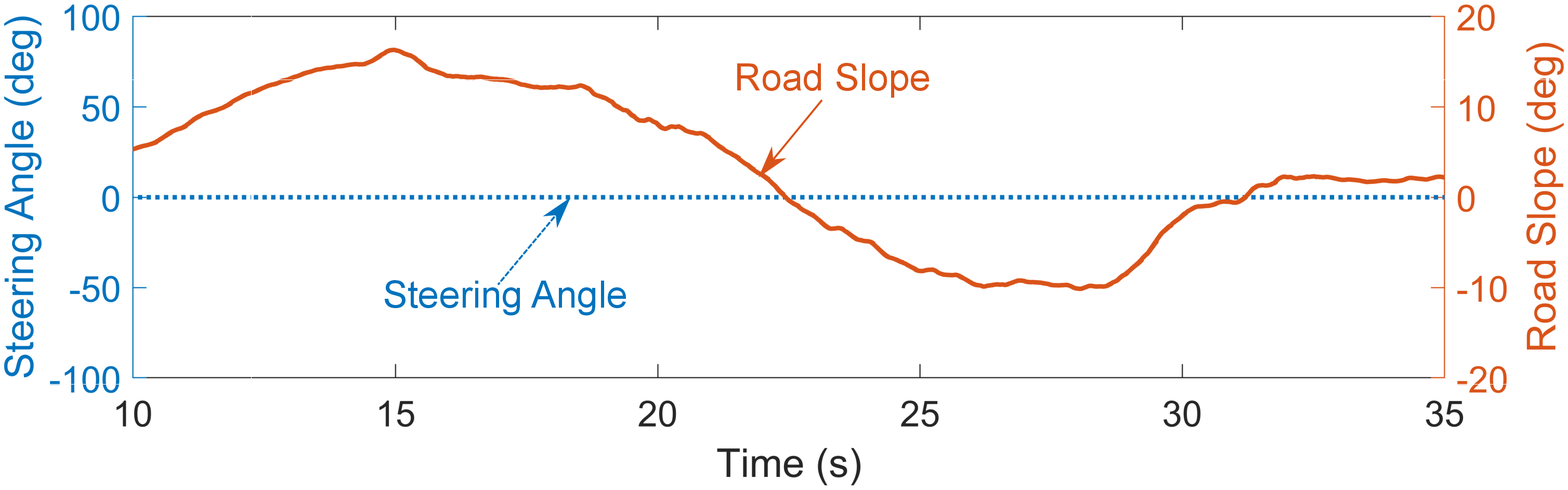}
\subcaption{}
    \label{5a}
\end{minipage}
\begin{minipage}[b]{\minipagetextwidth\textwidth}
		\includegraphics[width=\textwidth]{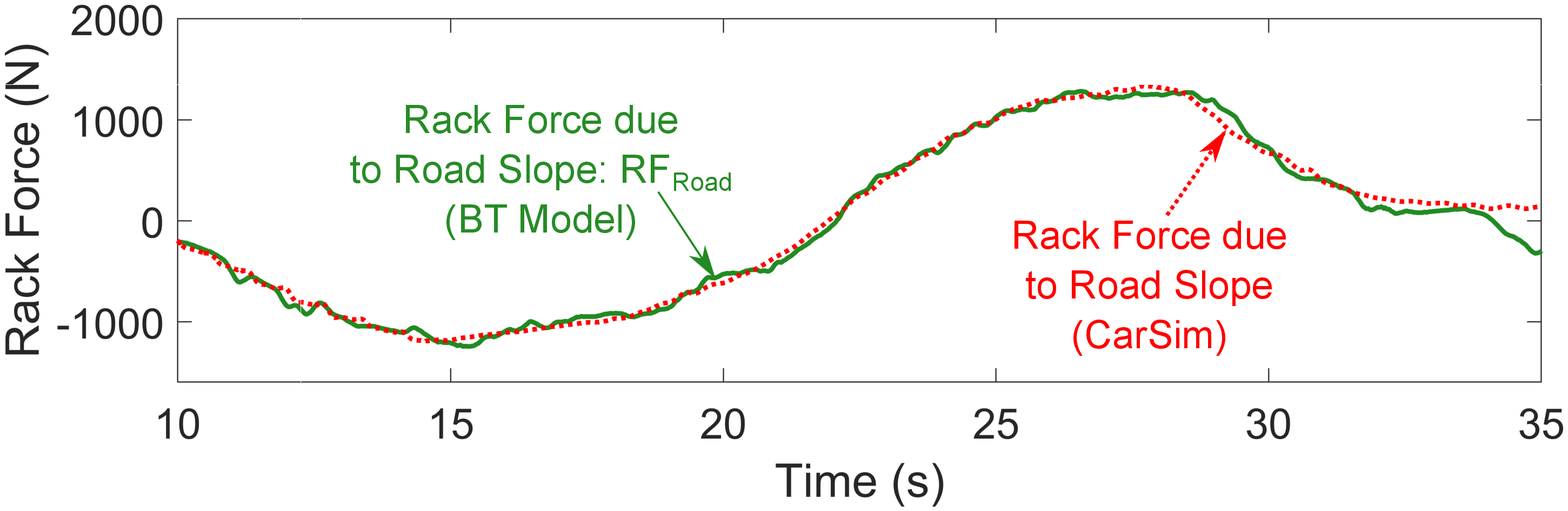}
\subcaption{}
    \label{5b}
\end{minipage}
	\begin{minipage}[b]{\minipagetextwidth\textwidth}
		\includegraphics[width=\textwidth]{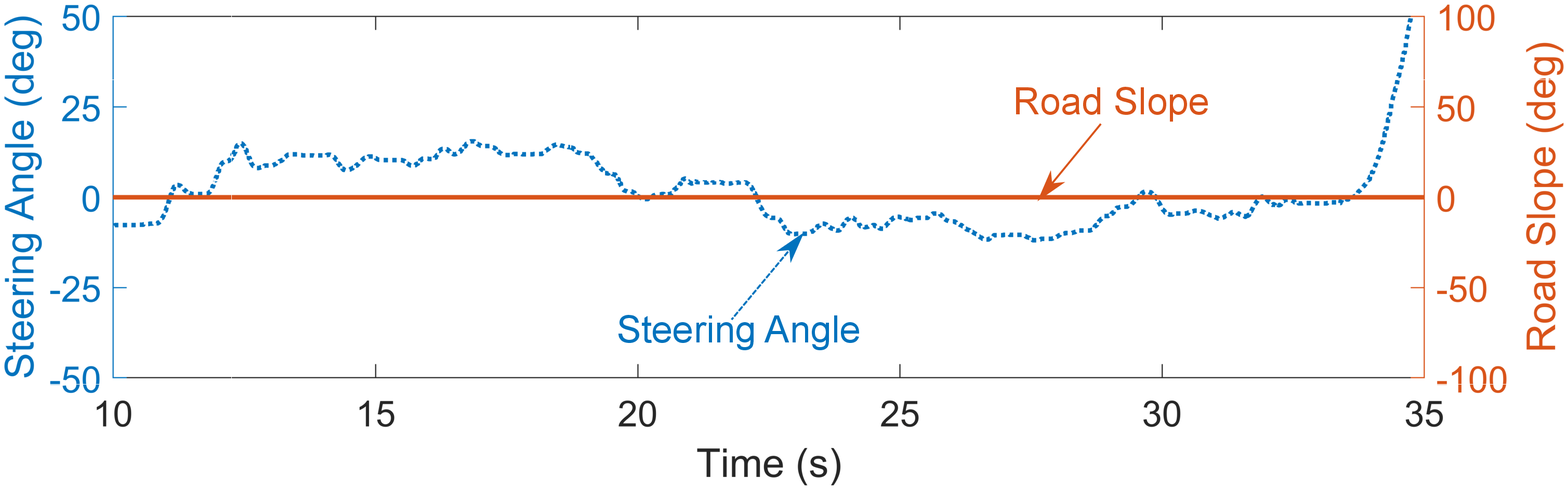}
 \subcaption{}
    \label{5c}
\end{minipage}
\begin{minipage}[b]{\minipagetextwidth\textwidth}
		\includegraphics[width=\textwidth]{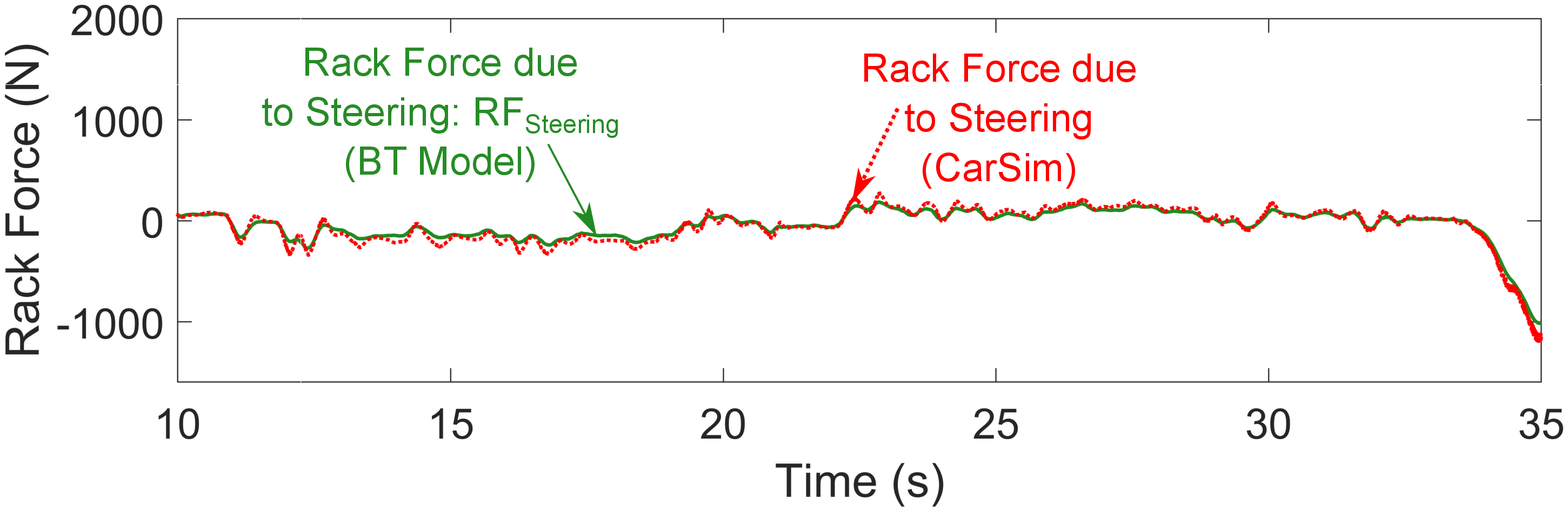}
 \subcaption{}
    \label{5d}
\end{minipage}
	\caption{Comparison of component wise estimates of rack force generated by the BT Model with the estimates generated by CarSim in Experiment 1. Vehicle speed maintained at about 20 km/h. (a) Input zero steering angle and non-zero road slope. (b) $RF_{Road}$ estimated using the BT Model and using CarSim. (c) Input non-zero steering angle and zero road slope. (d) $RF_{Steering}$ estimated using the BT Model and using CarSim.} 
		\label{e5}
\end{figure}

Results from the CarSim simulation study (Fig. \ref{e5}) illustrate the accuracy of rack force estimates due to steering angle and road profile produced by the BT Model. Inputs used to estimate rack force due to road profile ($RF_{Road}$) are shown in Fig. \ref{5a}, and to estimate rack force due to steering angle ($RF_{Steering}$) are shown in Fig. \ref{5c}. The $RF_{Steering}$ estimated using the BT Model agreed well with $RF_{Steering}$ estimated using CarSim (Fig. \ref{5b}). The normalized mean absolute error between the estimate produced by the BT Model and the estimate produced by CarSim was found to be only 3.62\%. Likewise, the estimates of $RF_{Road}$ produced using the BT Model also matched the estimates produced using CarSim with an estimation error of only 5.00\%. 

\begin{figure}[h!]
    \centering
\begin{minipage}[b]{\minipagetextwidth\textwidth}
		\includegraphics[width=\textwidth]{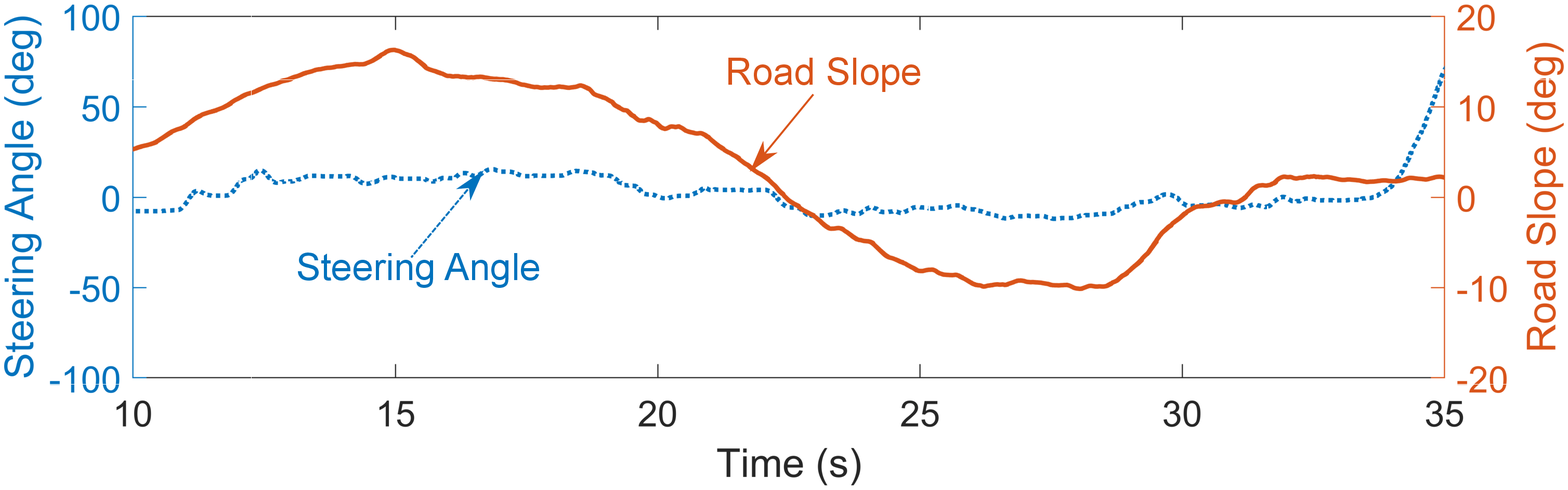}
\subcaption{}
    \label{6a}
\end{minipage}
\begin{minipage}[b]{\minipagetextwidth\textwidth}
		\includegraphics[width=\textwidth]{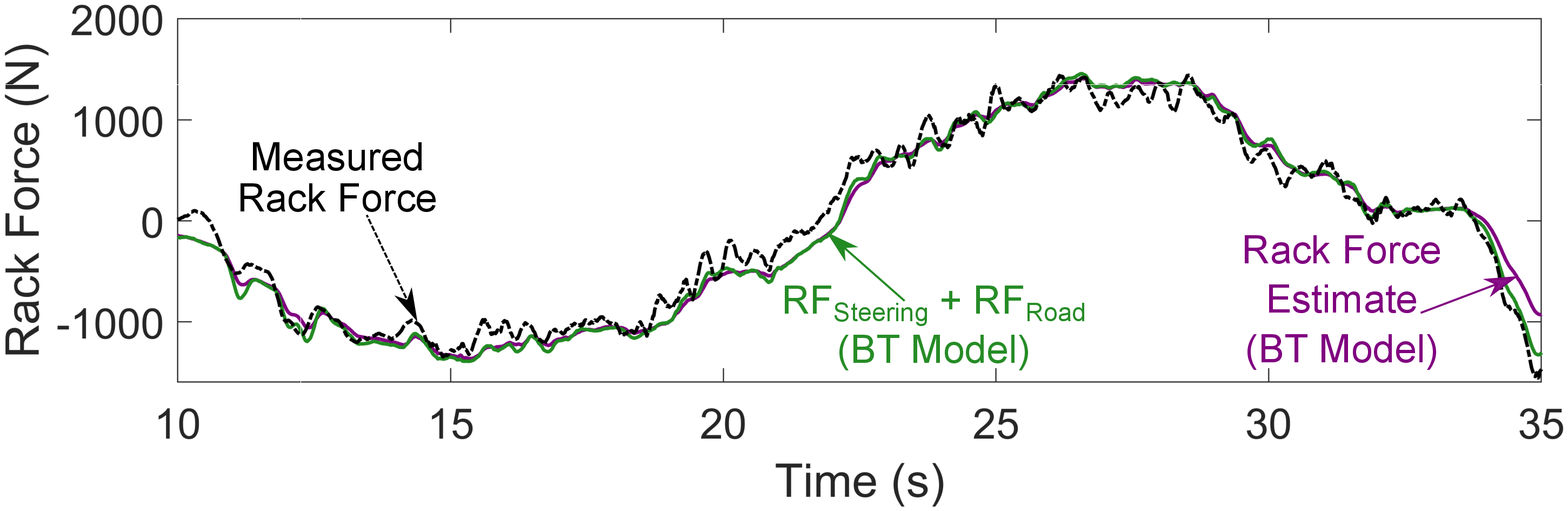}
\subcaption{}
    \label{6b}
\end{minipage}
	\caption{Comparison of total rack force estimate as a sum of component-wise rack force estimates generated by the BT Model with the measured rack force and the total rack force estimated using the BT Model. Vehicle speed maintained at about 20 km/h. (a) Input non-zero steering angle and non-zero road slope. (b) Sum of component-wise rack force estimates generated by the BT Model, total rack force generated by the BT Model, and rack force measured using sensor.} 
		\label{e6}
\end{figure}

Next, we investigated how much the residual rack force $\Delta RF$ contributes to the total rack force. To this end, we compared the sum of the component-wise estimates $RF_{Steering}$ and $RF_{Road}$ with the total steering rack force estimated by the BT model and to the rack force measured using the rack force sensor mounted in the vehicle (see Fig. \ref{e6}). The input steering angle and road slopes for this test were modeled after the original inputs to Experiment 1. The sum of component-wise estimates of rack force matched well with the total steering rack force estimated by the BT Model (Fig. \ref{6b}). The normalized mean absolute error between $RF_{Steering}+RF_{Road}$ and the rack force $RF$ estimated using the BT Model was found to be only 1.33\%. In other words, the influence of residual rack force $\Delta RF$ on total rack force was found to be negligible.

The results from the simulation study show that for driving on road slopes with large variations, the BT Model is capable of producing component-wise estimates of rack force to perform targeted compensation. Considering the higher DOF VTM-based estimator in CarSim as a reference, the independent estimates of rack force due to steering angle and due to road profile produced by the simpler BT Model were both found to be accurate. Moreover, at least for the road slope variation of about $-13^{\circ}$ to $13^{\circ}$ and steering angle variation of about $-20^{\circ}$ to $20^{\circ}$, the residual rack was found to be negligible and rack force turned out to be mostly composed of $RF_{Steering}$ and $RF_{Road}$.

The results on the component-wise estimates of rack force can be utilized in the design of power steering control algorithms and driver assist features. Using the independent estimates of the rack force components, controllers may be designed to enable targeted compensation for enhanced steering feel (as suggested in \cite{hackl2000method, yang2014new}). Rack force components can also be suppressed entirely to determine a fault or undesirable behavior in the steering system \cite{greul2012determining}. Targeted compensation may also have critical applications in the steer-by-wire systems, driving simulators, and hardware-in-loop simulators where re-creating road feel and tuning the steering feel have always been an active area of research \cite{nehaoua2012rack, segawa2006preliminary}. The individual components of rack force can also be selectively tuned and displayed to the driver to realize different driving modes such as ``luxurious" or ``sporty" driving as discussed in \cite{greul2012determining}. 

\section{Conclusions and Future Work}\label{sec:conclusions}

In this paper, we developed three vehicle and tire model based (VTM-based) rack force estimators that can be used to estimate rack force for driving on uneven roads. We used the estimators to investigate the level of tire model complexity required to accurately estimate rack force for driving on low and high frequency road profile variations with non-aggressive and aggressive steering maneuvers. We found that estimators with non-linear tire models produce more accurate rack force estimates for driving with aggressive steering maneuvers. Moreover, only the relatively complex semi-empirical tire models such as the Rigid Ring tire model can accurately capture the rack force variation for driving on high frequency road profile variations. In addition, we checked whether the estimates of rack force due to steering angle and due to road profile produced by a VTM-based estimator could be used to perform targeted compensation to enhance the steering feel. 
To this end, we showed that, for a driving maneuver with large road slope variation, the component-wise estimates of rack force produced by our estimator match the component-wise estimates produced by a higher DOF vehicle and tire model estimator available in a commercial vehicle dynamics package (CarSim). Through one experiment, we also showed that rack force seems to only consist of rack force due to steering angle and due to road profile even though rack force is a nonlinear function of steering angle and road profile. 

While our estimators produced accurate estimates of rack force at normal driving speeds, the estimation accuracy dropped during parking maneuvers (low speeds and large steering angles). Moreover, in the majority of driving experiments that we performed, the speed was maintained consistent in order to isolate the effect of steering angle and road profile on the rack force. Future work can focus on developing VTM-based estimators that support rack force estimation for parking maneuvers and focus on testing the effect of speed variations on the estimation and decomposition of rack force. In comparison to other existing estimation techniques, VTM-based estimators use more tire and vehicle specific parameters to estimate the rack force. Future studies can investigate robustness of VTM-based estimators against the uncertainty in the values of these parameters. Future work can also focus on using VTM-based estimators to support development and improvement of electric power steering control algorithms and advanced driver assist functions. The ability of a VTM-based rack force estimator to produce component-wise estimates of rack force due to steering angle and due to road profile can also be used to perform targeted compensation on each individual component to improve steering feel.

\section*{Disclosure statement}

No potential conflict of interest was reported by the authors.

\section*{Funding}

This work was supported by the Ford Motor Company under a Ford/U-M Alliance
Project UM0146.

\bibliographystyle{tfnlm}
\bibliography{VSD_BIB.bib}

\begin{thebibliography}{10}
\providecommand{\url}[1]{\normalfont{#1}}
\providecommand{\urlprefix}{Available from: }

\bibitem{strecker2014method}
Strecker~J, Werner~T. Method for determining a toothed rack force for a
  steering device in a vehicle ; 2014. US Patent 8,788,147.

\bibitem{yang2014new}
Yang~T. A new control framework of electric power steering system based on
  admittance control. IEEE Transactions on Control Systems Technology.
  2014;\hspace{0pt}23(2):762--769.

\bibitem{pick}
Pick~AJ, Sworn~TJ, Barton~AD, et~al. Rack force disturbance rejection ; 2007.
  US Patent 7,273,127.

\bibitem{gruner2008control}
Gr{\"u}ner~S, Gaedke~A, Karch~GG. Control of electric power steering
  systems-from state of art to future challenges. In: Proceedings of the 17th
  World Congress The International Federation of Automatic Control; 2008. p.
  10756--10757.

\bibitem{dornhege2017steering}
Dornhege~J, Nolden~S, Mayer~M. Steering torque disturbance rejection. SAE
  International Journal of Vehicle Dynamics, Stability, and NVH.
  2017;\hspace{0pt}1(2017-01-1482):165--172.

\bibitem{greul2012determining}
Greul~R, Werner~T. Determining a target steering torque in a steering device ;
  2012. US Patent 8,249,777.

\bibitem{fankem2014model}
Fankem~S, Weiskircher~T, M{\"u}ller~S. Model-based rack force estimation for
  electric power steering. IFAC Proceedings Volumes.
  2014;\hspace{0pt}47(3):8469--8474.

\bibitem{vinattieri2016target}
Vinattieri~F, Wright~T, Capitani~R, et~al. Target setting and structural design
  of an {EPS}-in-the-loop test bench for steering feeling simulation. SAE
  Technical Paper; 2016.

\bibitem{greul2016method}
Greul~R, Werner~T, Strecker~J. Method for determining a rack force for a
  steering apparatus and steering apparatus ; 2016. US Patent 9,272,732.

\bibitem{kezobo2014electric}
Kezobo~I, Kurishige~M, Endo~M, et~al. Electric power steering control device ;
  2014. US Patent 8,626,394.

\bibitem{nehaoua2012rack}
Nehaoua~L, Djemai~M, Pudlo~P. Rack force feedback for an electrical power
  steering simulator. In: Control \& Automation (MED), 2012 20th Mediterranean
  Conference on; IEEE; 2012. p. 79--84.

\bibitem{bajcinca2006road}
Bajcinca~N, Nuthong~C, Svaricek~F. Road feedback estimation for steer-by-wire
  control. In: 2006 IEEE Conference on Computer Aided Control System Design,
  2006 IEEE International Conference on Control Applications, 2006 IEEE
  International Symposium on Intelligent Control; IEEE; 2006. p. 1288--1293.

\bibitem{bhardwaj2019estimating}
Bhardwaj~A, Gillespie~B, Freudenberg~J. Estimating rack force due to road
  slopes for electric power steering systems. In: 2019 American Control
  Conference (ACC); IEEE; 2019. p. 328--334.

\bibitem{blommer2012systems}
Blommer~MA, Sanders~PG, Tseng~HE, et~al. Systems and methods for decoupling
  steering rack force disturbances in electric steering ; 2012. US Patent
  8,150,582.

\bibitem{wang2016epas}
Wang~D, Esser~F. {EPAS} system tests using rack force models. SAE Technical
  Paper; 2016.

\bibitem{tseng2001dynamic}
Tseng~HE. Dynamic estimation of road bank angle. Vehicle system dynamics.
  2001;\hspace{0pt}36(4-5):307--328.

\bibitem{eric2007estimation}
Eric~Tseng~H, Xu~L, Hrovat~D. Estimation of land vehicle roll and pitch angles.
  Vehicle System Dynamics. 2007;\hspace{0pt}45(5):433--443.

\bibitem{viner1995rollovers}
Viner~JG. Rollovers on sideslopes and ditches. Accident Analysis \& Prevention.
  1995;\hspace{0pt}27(4):483--491.

\bibitem{peters2006modeling}
Peters~SC. Modeling, analysis, and measurement of passenger vehicle stability
  [dissertation]. Massachusetts Institute of Technology; 2006.

\bibitem{weiskircher2015rack}
Weiskircher~T, Fankem~S, Ayalew~B. Rack force estimation for electric power
  steering. In: ASME 2015 International Design Engineering Technical
  Conferences and Computers and Information in Engineering Conference; American
  Society of Mechanical Engineers; 2015.

\bibitem{hackl2000method}
Hackl~M, Kraemer~W. Method and apparatus for operating a steering system for a
  motor vehicle ; 2000. US Patent 6,085,860.

\bibitem{karnopp1993motor}
Karnopp~D. Motor-driven servo steering system ; 1993. US Patent 5,205,371.

\bibitem{ikeda2011electric}
Ikeda~H, Endo~M, Kezobo~I. Electric power-steering control apparatus ; 2011. US
  Patent 8,050,825.

\bibitem{honisch2015improvement}
Honisch~A, Lugert~M, Sch{\"o}ning~T, et~al. Improvement of steering feel
  virtual approach with {HiL}. ATZ worldwide. 2015;\hspace{0pt}117(6):10--13.

\bibitem{toyohira2010validity}
Toyohira~T, Nakamura~K, Tanno~Y. The validity of {EPS} control system
  development using {HIL}s. SAE Technical Paper; 2010.

\bibitem{marino2010asymptotic}
Marino~R, Scalzi~S. Asymptotic sideslip angle and yaw rate decoupling control
  in four-wheel steering vehicles. Vehicle System Dynamics.
  2010;\hspace{0pt}48(9):999--1019.

\bibitem{gadola2014development}
Gadola~M, Chindamo~D, Romano~M, et~al. Development and validation of a kalman
  filter-based model for vehicle slip angle estimation. Vehicle System
  Dynamics. 2014;\hspace{0pt}52(1):68--84.

\bibitem{koch2010untersuchungen}
Koch~T. Untersuchungen zum {L}enkgef{\"u}hl von steer-by-wire {L}enksystemen
  [dissertation]. Technische Universit{\"a}t M{\"u}nchen; 2010.

\bibitem{segawa2006preliminary}
Segawa~M, Nakano~S, Shino~M, et~al. Preliminary study concerning quantitative
  analysis of steering system using hardware-in-the-loop ({HIL}) simulator. SAE
  Technical Paper; 2006.

\bibitem{bhardwaj2020rack}
{Bhardwaj}~A, {Slavin}~D, {Walsh}~J, et~al. Rack force estimation for driving
  on uneven road surfaces. arXiv preprint; arXiv:2002.01655.  2020.

\bibitem{pastor1995method}
Pastor~SR, Tierney~GL. Method and apparatus for estimating incline and bank
  angles of a road surface ; 1995. US Patent 5,446,658.

\bibitem{rajamani2011vehicle}
Rajamani~R. Vehicle dynamics and control. Springer Science \& Business Media;
  2011.

\bibitem{pacejka2005tire}
Pacejka~H. Tire and vehicle dynamics. Elsevier; 2005.

\bibitem{schmeitz2004semi}
Schmeitz~AJC. A semi-empirical three-dimensional model of the pneumatic tyre
  rolling over arbitrarily uneven road surfaces [dissertation]; 2004.

\bibitem{balachandran2014virtual}
Balachandran~A, Erlien~SM, Gerdes~JC. The virtual wheel concept for supportive
  steering feedback during active steering interventions. In: ASME 2014 Dynamic
  Systems and Control Conference; American Society of Mechanical Engineers
  Digital Collection; 2014.

\bibitem{zegelaar1996plane}
Zegelaar~P, Pacejka~H. The in-plane dynamics of tyres on uneven roads. Vehicle
  System Dynamics. 1996;\hspace{0pt}25(S1):714--730.

\bibitem{carsim2017math}
Carsim math models. Mechanical Simulation Corporation, www.carsim.com; 2017.

\bibitem{bakker1989new}
Bakker~E, Pacejka~HB, Lidner~L. A new tire model with an application in vehicle
  dynamics studies. SAE transactions. 1989;\hspace{0pt}98(6):101--113.

\bibitem{pacejka1991shear}
Pacejka~HB, Sharp~RS. Shear force development by pneumatic tyres in steady
  state conditions: a review of modelling aspects. Vehicle system dynamics.
  1991;\hspace{0pt}20(3-4):121--175.

\bibitem{sayers2018tire}
Sayers~M. Tire models. Mechanical Simulation Corporation, www.carsim.com; 2018.

\bibitem{raad2014pull}
Raad~JM, Tuttle~D. Pull-drift compensation enhancements ; 2014. US Patent
  8,798,865.

\bibitem{shah2011pull}
Shah~J. Pull drift compensation using active front steering system ; 2011. US
  Patent App. 12/848,242.

\bibitem{bosch2002tyre}
Bosch~P, Ammon~D, Klempau~F. Tyre models--desire and reality in respect of
  vehicle development. Darmstadter Reifenkolloquium.
  2002;\hspace{0pt}17:87--101.

\bibitem{izadi2020impedance}
{Izadi}~V, {Bhardwaj}~A, {Ghasemi}~AH. Impedance modulation for negotiating
  control authority in a haptic shared control paradigm. arXiv preprint;
  arXiv:2001.07779.  2020.

\bibitem{gordon2015automated}
Gordon~T, Lidberg~M. Automated driving and autonomous functions on road
  vehicles. Vehicle System Dynamics. 2015;\hspace{0pt}53(7):958--994.

\end{thebibliography}

\end{document}